\documentclass{emulateapj}

\usepackage{amsmath}
\usepackage{amssymb}
\usepackage{graphicx}

\newcommand{\MCcode}{{\sc sedona}}
\newcommand{\kms}{\ensuremath{\mathrm{km~s}^{-1}}}
\newcommand{\Nifs}{\ensuremath{^{56}\mathrm{Ni}}}
\newcommand{\Nife}{\ensuremath{^{58}\mathrm{Ni}}}
\newcommand{\Cofs}{\ensuremath{^{56}\mathrm{Co}}}

\newcommand{\Feff}{\ensuremath{^{54}\mathrm{Fe}}}
\newcommand{\msun}{{\ensuremath{\mathrm{M}_{\odot}}}}
\newcommand{\Msun}{{\ensuremath{\mathrm{M}_{\odot}}}}
\newcommand{\zsun}{\ensuremath{Z_\odot}}
\newcommand{\texp}{\ensuremath{t_{\mathrm{exp}}}}
\newcommand{\Mni}{\ensuremath{M_{\mathrm{Ni}}}}
\newcommand{\Mfe}{\ensuremath{M_{\mathrm{Fe}}}}
\newcommand{\Mime}{\ensuremath{M_{\mathrm{IME}}}}
\newcommand{\Mburn}{{\ensuremath{\mathrm{M}_{\mathrm{burn}}}}}
\newcommand{\NiCent}{\ensuremath{M^c_{\mathrm{Ni}}}}
\newcommand{\Mch}{{\ensuremath{\mathrm{M}_\mathrm{ch}}}}
\newcommand{\lSect}[1]{{\label{sec:#1}}}

\newcommand{\KE}{\ensuremath{E_K}}
\newcommand{\Mb}{\ensuremath{M_B}}

\newcommand{\dmfb}{\ensuremath{\Delta M_{15}(B)}}

\newcommand{\phiNi}{{\ensuremath{\phi_\mathrm{Ni}}}}
\newcommand{\epth}{{\ensuremath{\epsilon_{\rm th}}}}

\def\gtaprx {\lower .1ex\hbox{\rlap{\raise .6ex\hbox{\hskip .3ex
	{\ifmmode{\scriptscriptstyle >}\else
		{$\scriptscriptstyle >$}\fi}}}
	\kern -.4ex{\ifmmode{\scriptscriptstyle \sim}\else
		{$\scriptscriptstyle\sim$}\fi}}}
\def\ltaprx {\lower .1ex\hbox{\rlap{\raise .6ex\hbox{\hskip .3ex
	{\ifmmode{\scriptscriptstyle <}\else
		{$\scriptscriptstyle <$}\fi}}}
	\kern -.4ex{\ifmmode{\scriptscriptstyle \sim}\else
		{$\scriptscriptstyle\sim$}\fi}}}

\newcommand{\Sectff}[1]{{\ref{sec:#1}}}
\newcommand{\Sect}[1]{{\S~\Sectff{#1}}}

\def\ergg{erg~g$^{-1}$}

\begin{document}

\shorttitle{Type Ia Supernova Light Curves}
\shortauthors{Woosley et al.}

\title{Type~Ia Supernova Light Curves}

\author{S. E. Woosley\altaffilmark{1}, D. Kasen\altaffilmark{2,3},
S. Blinnikov\altaffilmark{4,5}, and E. Sorokina\altaffilmark{4}}

\altaffiltext{1}{Department of Astronomy and Astrophysics, University
of California, Santa Cruz, CA 95064; woosley@ucolick.org}
\altaffiltext{2}{Department of Physics and Astronomy, Johns Hopkins
University, Baltimore, MD, 21218} 
\altaffiltext{3}{Space Telescope Science Institute, Baltimore, MD, 21218} 
\altaffiltext{4}{Institute for Theoretical and Experimental Physics,
Moscow} 
\altaffiltext{5}{Max Planck Institut f\"ur Astrophysik, Garching,
Germany}

\begin{abstract} 
The diversity of Type Ia supernova (SN Ia) photometry is explored
using a grid of 130 one-dimensional models. It is shown that the
observable properties of SNe~Ia resulting from Chandrasekhar-mass
explosions are chiefly determined by their final composition and some
measure of ``mixing'' in the explosion. A grid of final compositions
is explored including essentially all combinations of $^{56}$Ni,
stable ``iron'', and intermediate mass elements that result in an
unbound white dwarf. Light curves (and in some cases spectra) are
calculated for each model using two different approaches to the
radiation transport problem. Within the resulting templates are models
that provide good photometric matches to essentially the entire range
of observed SNe~Ia.  On the whole, the grid of models spans a wide
range in $B$-band peak magnitudes and decline rates, and does not obey
a Phillips relation. In particular, models with the same mass of
$^{56}$Ni show large variations in their light curve decline rates.
We identify and quantify the additional physical parameters
responsible for this dispersion, and consider physically motivated
``cuts'' of the models that agree better with the Phillips relation,
discussing why nature may have preferred these solutions. For example,
models that produce a constant total mass of burned material of 1.1
$\pm 0.1$ \msun\ do give a crude Phillips relation, albeit with much
scatter. If one further restricts that set to models that make 0.1 to
0.3 \msun\ of stable iron and nickel isotopes, and then mix the ejecta
strongly between the center and 0.8 \msun, reasonable agreement with
the Phillips relation results, though still with considerable
spread. We conclude that the supernovae that occur most frequently in
nature are highly constrained by the Phillips relation and that a
large part of the currently observed scatter in the relation is likely
a consequence of the intrinsic diversity of these objects.
\end{abstract}


\section{INTRODUCTION}
\lSect{intro}

The application of Type Ia supernovae (SNe~Ia) to cosmological
distance determination has yielded revolutionary insights into the
structure and composition of the universe \citep{Per99,Rie98}. The
utility of such explosions is based upon two empirical
observations. First, that Type Ia supernovae are, for the most part,
approximate standard candles, even without any corrections being
applied. This probably reflects their common origin in the explosion
of a white dwarf of standard mass \citep[see review by][]
{Hil99}. Second, a large part of the residual diversity in peak
brightness can be corrected for by use of either light curve template
fitting or observed correlations between light curve decline rate and
peak brightness - the so called ``Phillips Relation''\citep{Phi93}, or
``width-luminosity relation'' (henceforth WLR)\footnote[1]{It should
  be noted that a correlation of the peak luminosity and the decline
  rate - with the correct sign - was already established by
  \citet{Psk77}. See the history and other references in
  \citet{Phi05}.}.

While the empirical relation between brightness and width has worked
well for most purposes, as we move into an era of ``precision
cosmology'', one must feel increasing unease at the lack of an agreed
upon standard model for how Type Ia supernovae explode and the
possibility that evolutionary or environmental factors may erode
accuracy at large distances. If the mass of $^{56}$Ni made in a
supernova is the dominant physical parameter affecting both brightness
and light curve shape, what are the magnitude and direction of other
possible parameters of the explosion such as kinetic energy,
intermediate mass element production, stable iron production and
mixing?  And if, as we shall find, not all physically plausible models
obey a WLR, why has nature chosen to realize frequently a particular
subset? That is, what does the WLR tell us, not just about cosmology,
but about supernova models? If one can make progress on both these
questions, then it may be possible to derive more rigorous and
quantitative tools for using Type Ia supernovae for distance
determination. That is the long term goal of our investigation.

In this paper, we introduce a simple parameterized approach to
computing SN~Ia explosions and use it to generate
a large grid of one-dimensional models.  For each model, we calculate
synthetic broadband light curves and examine the relationship between
peak $B$ magnitude and $B$-band decline rate.  Such an approach allows
us to study the physical parameters affecting the light curves of
SNe~Ia and the WLR.  Several previous theoretical studies have
performed similar investigations for a given set of models
\citep{Khokhlov_93,Hoe96a,Hoe98,Pin01, Hoeflich_99by,Mazzali_WLR}.  These
studies, however, were typically confined to a certain class (or
classes) of theoretical explosion paradigms.  In this study, rather
than adopt a specific theoretical framework, we take a general and
exhaustive approach by simulating the full range of final ejecta
structures conceivably arising from the disruption of a
Chandrasekhar-mass carbon-oxygen white dwarf.

\section{A SIMPLE MODEL FOR THE EXPLOSION}

\subsection{Construction and Parameters of the Model}
\lSect{toys}

Here a simple {\sl ansatz} for the explosion model is introduced that
turns out to work surprisingly well. It is assumed that all SNe~Ia
start from a common point, a 1.38 \msun\ carbon oxygen white dwarf
with a central density $2 - 3 \times 10^9$ g cm$^{-3}$. Burning, which
may be quite complicated, turns some part of that fuel into ash and
deposits an internal energy in the dwarf equal to the difference in
nuclear binding energy between the initial and final
compositions. This composition is imprinted on the white dwarf and, in
our calculations, the corresponding change in nuclear binding energy
is deposited uniformly throughout its mass.  The white dwarf expands
to infinity with a velocity distribution determined by that energy.
The $^{56}$Ni and $^{56}$Co decay, and the time-dependent radiation
transport is calculated.

The major assumption that facilitates the calculation is that the
energy deposition is shared globally. This is clearly true in
spherically symmetric deflagration models, since sound waves move much
faster than the flame and share the overpressure created by the
burning with the rest of the star. It is also true in detonations
except for a thin layer near the surface. Throughout most of the star,
the passage of a detonation wave changes the composition and internal
energy. Most acceleration occurs afterwards. After expansion of a
factor of one million (before it can be seen), and the development of
a velocity field that must increase with Lagrangian mass, the results
are the same as achieved by a simple composition swap and artificial
energy deposition.  Perhaps the best validation of the model is that
it works. Calculations given in \Sect{W7} and \Sect{compobs} show that
the multi-band photometry of models calculated this way agrees, both
with previous detailed numerical simulations of the explosion, like
the well-studied Model W7, and with a diverse set of observed
supernovae.

However, the simple model has one basic shortcoming.  It will turn out
that ``mixing'' - just how a given composition is distributed in
velocity - is an important parameter of the problem. It is this
information that must eventually be provided by a full,
first-principle's calculation. Buried within this mixing parameter is
also information about the possible asymmetry of the explosion.  Such
simulations must ultimately be three dimensional. However, the present
models are as ``physical'' as most other one-dimensional
approximations that make assumptions about flame speeds, transitions
to detonation, metallicity, etc.

The dominant products of burning are assumed to be: i) $^{56}$Ni; ii)
``stable iron'', that is all the other nuclei in the iron group,
chiefly $^{54}$Fe and $^{58}$Ni; and iii) intermediate mass elements -
Si, S, Ar, Ca, henceforth referred to as ``IME''. In the latter group
Si and S are most abundant and, for making energy, most important, but
Ca is important for the spectrum. These are the parameters of the
solution. Here IME ratios are adopted from Model DD4 of \citet{Woo94}:
by mass Si (53\%), S (32\%), Ar (6.2 \%) and Ca (8.3\%).  Solar ratios
\citep{Lod03} would not have been much different: Si (58\%), S (29\%),
Ar (7.7\%), and Ca (5.2\%).  There is some physical motivation for
this as well. At the temperatures where carbon and oxygen burn to
silicon, 3 - 5 $\times 10^9$ K, quasi-equilibrium favors an
approximately solar abundance set \citep{Woo73}. Stable iron is taken
in the calculation to be $^{54}$Fe. A further refinement in which
stable iron is split into $^{54}$Fe and $^{58}$Ni could easily be
done, but would add an additional parameter without greatly affecting
the light curve.

The relative amounts of these three sets of burning products reflect
specific physical processes in the star and play a unique role in
making the light curve and spectrum. ``Stable iron'' is a measure of
burning at temperatures and densities so high (T $\gtaprx 5 \times
10^9$ K; $\rho \gtaprx 10^8$ g cm$^{-3}$) that nuclear statistical
equilibrium is attained and accompanied by electron capture. To some
extent, stable iron is also a function of the initial metallicity of
the star \citep{Tim03}. Stable iron-group isotopes contribute to the
opacity and explosion energy, but not to the later energy generation
that makes the light curve. A higher ignition density, which might
reflect a lower accretion rate, increases the production of stable
iron. The most natural location for stable iron is in the center of
the supernova where the density was the highest, but iron will also
exist at a level of about 5\% (Z/\zsun) by mass in the $^{56}$Ni
layer.  This is a consequence of the $^{22}$Ne present in the initial
composition from helium burning, and the number is sensitive
(linearly) to the metallicity.

The double magic nucleus $^{56}$Ni is always the dominant product of
nucleosynthesis starting from a fuel with equal numbers of neutrons
and protons (like $^{12}$C and $^{16}$O) when nuclear statistical
equilibrium is attained. Its abundance therefore reflects the extent
of nuclear burning above $5 \times 10^9$ K at densities low enough
that electron capture is negligible. Without doubt, it is the key
player in the SN~Ia light curve. Along with its decay product
$^{56}$Co, it powers the light curve and also contributes to the
opacity and explosion energy.  The most natural location for the
$^{56}$Ni is also deep inside the star just outside the stable
iron. However, turbulence, instabilities in the explosion, and perhaps
delayed detonation may lead to some of the $^{56}$Ni (and to some
extent, stable iron) being ``mixed'' far out, perhaps even to the
surface.

IME are made when the density in the burning region declines to about
10$^7$ g cm$^{-3}$. There, the heat capacity of the radiation field
keeps the burning from going all the way to equilibrium. While these
elements cause prominent line features in the SN~Ia spectrum, their
opacities and emissivities are not as important to the light curve as are
those of nickel, cobalt and iron.  Their production does, however,
contribute to the explosion energy.  The most natural location of IME
is in the outer layers of the supernova, although instabilities and
delayed detonation can lead to mixing among the layers in velocity
space.

Given an initial white dwarf, the free parameters of the calculation
are thus \Mni, \Mfe, and \Mime. To this list one must also add
``mixing'', which is harder to quantify. The carbon-to-oxygen ratio
and the white dwarf binding energy (or equivalently ignition density)
also affect the overall energetics of the explosion. Burning carbon
releases a little more energy than burning oxygen and a more tightly
bound white dwarf requires more energy to give a certain expansion
speed. Here we assume a standard, near Chandrasekhar-mass white dwarf
(1.38 \msun) composed of equal amounts of carbon and oxygen (50\% by
mass fraction of each) with an ignition density of $2.9 \times 10^9$ g
cm$^{-3}$. This corresponds to a net binding energy (internal energy
plus gravitational binding energy) of $4.62 \times 10^{50}$ erg
(Coulomb corrections are neglected in this study). The ignition
density cannot be much greater than this or the production of rare
neutron-rich species will exceed the Galactic inventory as represented
by the sun \citep{Woo97}. The runaway will not ignite at densities
below about $2.5 \times 10^9$ g cm$^{-3}$ for reasonable choices of
the electron screening function.

The energy of the explosion is determined by the energy released in
the nuclear burning.  The asymptotic kinetic energy of the SN~Ia
explosion is directly related to the amount of material burned in the
explosion, in particular, for a starting composition of 50\% C and
50\% O,
\begin{equation}
\KE = 1.56 \Mni + 1.74 \Mfe + 1.24 \Mime - E_g + E_{int}
\end{equation}
where $E_g = 3.35$ B (here 1 B = 1 Bethe = 10$^{51}$ erg) is the
gravitational binding energy and $E_{int} = 2.89$ B, the internal
energy of the progenitor white dwarf.  If the composition were 30\% C
and 70\% O, the yields would be about 7\% less for Ni and Fe and 9\%
less for Si - Ca. A similar change in the opposite direction occurs
for an oxygen-rich initial composition - 70\% O and 30\% C.

This energy is deposited uniformly (as a certain number of erg
g$^{-1}$) throughout the entire mass of the initial white dwarf and
the composition of the layers changed appropriately (e.g.,
Fig.~\ref{Fig:mixing} for an explosion that made 0.7 \msun \ of $^{56}$Ni,
0.1 \msun \ of $^{54}$Fe and 0.3 \msun \ of Si-Ca), The ensuing
evolution is followed using the {\sc kepler} implicit hydrodynamics
code \citep{Wea78,Woo02}. After less than a minute (star time), the
explosion is homologously coasting and can be mapped into either of
the radiation codes discussed in \Sect{danscode} or \Sect{stella}. The
hydrodynamical calculation also takes less than one minute on a desktop
processor, thus allowing a large number of models with carefully
controlled properties to be generated.

\subsection{Mixing}
\lSect{mixing}

\begin{figure}
\begin{center}
\includegraphics[clip=true,width=8cm]{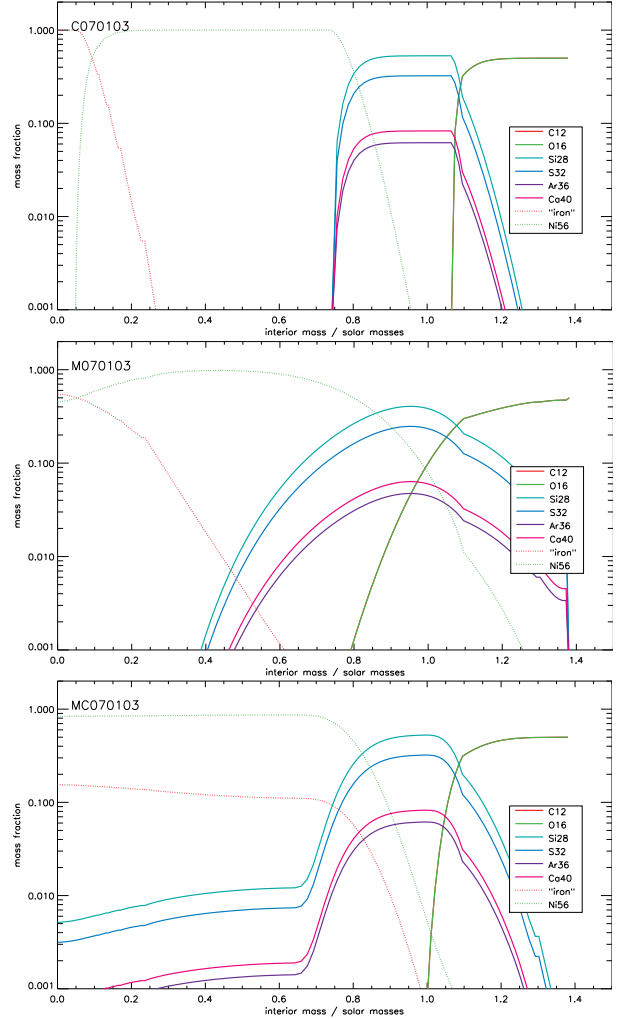}
\caption{Various parameterized mixing prescriptions applied to the
  model with 0.7 \Msun \ of \Nifs, 0.1 \Msun \ of stable iron, and 0.3
  \Msun \ of IME. The remainder is carbon and oxygen in equal (50\%)
  amounts by mass. The lines for C and O lie on top of each other in
  the figure. Model C070101 is very mildly mixed corresponding to
  moving boxcar average with a window of 0.02 \Msun \ applied three
  times. In Model M070103, the same star and composition has been
  mixed with the same 0.02 \Msun \ interval 50 times.  Model MC070103
  is extensively mixed inside 0.8 \Msun \ and mildly mixed outside
  (see text).
\label{Fig:mixing}}
\end{center}
\end{figure}

``Mixing'' is a general term referring to the fact that the ejected
composition is not stratified into shells of pure ashes from a given
burning process - here $^{54}$Fe, $^{56}$Ni, Si-Ca, and CO - but has
somehow become blended in velocity space. Mixing occurs at some level
simply because the burning temperature is continuous. Carbon and
oxygen don't burn only to Si-Ca or $^{56}$Ni, but, at some temperatures,
to a mixture of both. Stable iron may be present in the $^{56}$Ni
layer if the star had an appreciable (especially super-solar)
metallicity.  The most difficult mixing to quantify however, is a
consequence of the inherently multi-dimensional nature of the burning
which, from start to finish, is Rayleigh-Taylor unstable and
turbulent. Particular models like gravitationally confined detonation
\citep{Ple04,Roe06} even have most of the burning starting at the surface
and moving inwards.

A great variety of mixing is, in principle, possible and to the extent
that our results turn out to be sensitive to mixing, the simple ansatz
employed for the explosion is questionable. However, there are some
general rules that mixing should obey. First, mass and energy are
conserved, so the dynamics of the explosion, including the density
profile in the coasting configuration is probably not affected very
much. Second, the {\sl angle-averaged} atomic weight of the ejecta
probably decreases from the middle outwards. That is $^{56}$Ni and
stable iron are concentrated more towards the center, carbon and
oxygen on the outside, and Si-Ca, in between. Third, the neutron
excess, $\eta = \Sigma (N - Z) X/A$, also decreases from the center to
the surface.

A one-dimensional treatment of mixing may not be so bad, as far as the
radiative transfer is concerned.  The observed light curve is the
emission integrated over the entire star. Angle-dependent mixing can
certainly have consequences for individual spectral lines, but has a
smaller effect on the photometry.  This motivates the treatment of
mixing as a parameterized one-dimensional process that stirs, but does
not homogenize the composition. Here, that is accomplished by a
running boxcar average of abundances in a certain mass interval moved
from the center to surface of the star. In a typical mixing operation,
starting with zone 1, the composition in the next 0.02 \msun\ outwards
is homogenized. One then moves to zone 2 and does the same operation,
and so on until the stellar surface is reached. If more mixing is
desired, then either the mass interval is widened or the operation
repeated, from center to surface, multiple times. The latter approach
was adopted here. Mildly and moderately mixed versions of a sample
model are shown in Fig.~\ref{Fig:mixing}.

\subsection{Model Nomenclature}
\lSect{names}

Models are named here according to their composition and the mixing
prescription employed. Model Cxxyyzz is a ``mildly mixed'' model with
xx/10 solar masses of $^{56}$Ni, yy/10 solar masses of stable iron,
zz/10 solar masses of IME (Si-Ca), and 1.38 - (xx+yy+zz)/10 solar
masses of carbon and oxygen. Prior to mixing, unless otherwise
mentioned, the ordering of the composition is stable iron (center),
$^{56}$Ni next, and (Si-Ca) farthest out, followed by unburned carbon
and oxygen.  By mild mixing (Fig.~\ref{Fig:mixing}) we mean
specifically that a moving interval of mass of 0.02~\msun\ was mixed
from center to surface three times. ``Moderately mixed'' models,
Mxxyyzz, which are most frequently employed in this paper, apply the
same moving boxcar average 50 times. ``Highly mixed'' models,
MMxxyyzz, apply the same moving boxcar average 200 times.

A finer grid of the M-series models was also prepared for highlighting
a restricted range of burned mass (\Sect{constfe}), 1.0 and 1.1 \msun;
\Nifs\ masses, 0.4 to 0.9 \msun\; and stable iron masses, 0 - 0.3
\msun\ and a finer sampling of the intervals was also used.
Additional models were also constructed to explore different mixing
prescriptions described in \Sect{stablefe} and
\Sect{limmix}. Figure~\ref{Fig:mixing} shows the compositional
structure of a representative model with several mixing prescriptions.

\section{RADIATION TRANSPORT CALCULATIONS}
\lSect{transport}

The radiative transfer calculations, which by far consumed most of the
computer time, were performed using two independent radiative transfer
codes which employ very different numerical methods and atomic data.
Both start with a supernova that has already expanded to the
``coasting'' or ``homologous expansion'' phase. Operationally, that
means the explosion model from {\sc kepler} was linked into the
radiation transport code at an age of 10,000 s.

\subsection{Monte Carlo - Sedona}
\lSect{danscode}

The \MCcode\ code \citep{Kasen_Sedona} is a time-dependent
multi-dimensional Monte Carlo radiative transfer code, designed to
calculate the light curves, spectra and polarization of supernova
explosion models.  Given a homologously expanding SN ejecta structure,
\MCcode\ calculates the full time series of emergent spectra at high
wavelength resolution.  Broadband light curves are then constructed by
convolving the synthetic spectrum at each time with the appropriate
Bessel filter transmission functions \citep{Bessell_1990}. \MCcode\
includes a detailed treatment of gamma-ray transfer to determine the
instantaneous energy deposition rate from radioactive \Nifs\ and
\Cofs\ decay.  Radiative heating and cooling rates are evaluated from
Monte Carlo estimators, and the temperature structure of the ejecta
determined by iterating the model to thermal equilibrium.   See
\cite{Kasen_Sedona} for a detailed code description and verification.

Several significant approximations are made in \MCcode, notably the
assumption of local thermodynamic equilibrium (LTE) in computing the
atomic level populations.  In addition, bound-bound line transitions
are treated using the expansion opacity formalism (implying the
Sobolev approximation).  Although the \MCcode\ code is capable of a
direct Monte Carlo treatment of NLTE line processes, due to
computational constraints this functionality is not exploited here.
Instead, the line source functions are treated using an approximate
two-level atom approach (see Eq.~\ref{Eq:S}, next section).  In the
present calculations, we assume for simplicity that all lines are
``purely absorptive'', i.e., in the two-level atom formalism the ratio
of the probability of redistribution to pure scattering is taken to be
$\epsilon_{\rm th} = 1$ for all lines.  In this case, the line source
functions are given by the Planck function, consistent with our
adoption LTE level populations.  The one exception is the calcium
lines, which are assumed to be pure scattering ($\epsilon_{\rm th} =
0$) for the reasons discussed in
\cite{Kasen_IRLC}.

The assumption of purely absorbing lines has been common in previous
SN transfer calculations \citep{Pin01,Bli06}.  \cite{Kasen_Sedona}
demonstrate that the $\epth = 1$ approach does in fact capture the
true NLTE line fluorescence processes operative in SNe~Ia light curves,
however quantitative errors in the broadband magnitudes are expected
on the order of 0.1 to 0.3~mag.  A refined calibration would better
represent the true NLTE redistribution probabilities and may provide
more accurate results \citep[e.g.,][]{Hoe95}.  In particular,
\cite{Kasen_Sedona} shows that for SN~Ia models
assuming $\epth < 1$ leads to more accurate model colors in the
maximum light spectrum.  For this reason, our model peak magnitudes,
colors and decline rates should be considered uncertain by this
moderate amount.  The adoption of an alternative value $\epth \ne 1$
would lead to a shift in the location of the models in the WLR and
color plots discussed below, however this shift is most likely in a
uncorrelated way that does not significantly change the slope of the
model relation or the level of dispersion.  On the other hand, if
\epth\ depends in a systematic way on temperature or density (as to
some extent it must) this could in principle lead to correlated errors
that affect the slope of the model relation.  The issue will be the
subject of future investigations.

The two-level atom framework applied here is just one of several
uncertainties that affect the radiative transfer calculations.  In
addition, inaccuracy or incompleteness in the atomic line data can be
a source of significant error (see \Sect{codecomp}).  The inaccuracy
of the LTE ionization assumption may also have significant
consequences for the B-band light curves \citep{Kas06}. At later times
($\ga 30$~days after $B$-band maximum) the NLTE effects become
increasingly significant and the model calculations become unreliable.
One should keep in mind all the sources of uncertainty when
considering the model relations discussed below.

The numerical griding in the present calculations was as follows:
\emph{spatial:} 120 equally spaced radial zones with a maximum 
velocity of 30,000~\kms; \emph{temporal:} 100 time points beginning at
day~2 and extending to day 80 with logarithmic spacing $\Delta \ log\,
t = 0.175$;
\emph{wavelength:} covering the range 100-30000~\AA\ with resolution of
10~\AA.  Extensive testing confirms the adequacy of this griding for
the problem at hand. Atomic line list data was taken from the Kurucz
CD~1 line list
\citep{Kurucz_Lines}, which contains nearly 42 million lines.

A total of $10^7$ photon packets were used for each calculation, which
allowed for acceptable signal-to-noise (S/N) in the synthetic
broadband light curves.  For a few models, further time-independent
calculations were performed using additional photon packets in order
to calculate high S/N synthetic spectra at select epochs.

\subsection{Multi-energy group diffusion - Stella}
\lSect{stella}

\begin{figure}
\begin{center}
\includegraphics[clip=true,width=\columnwidth]{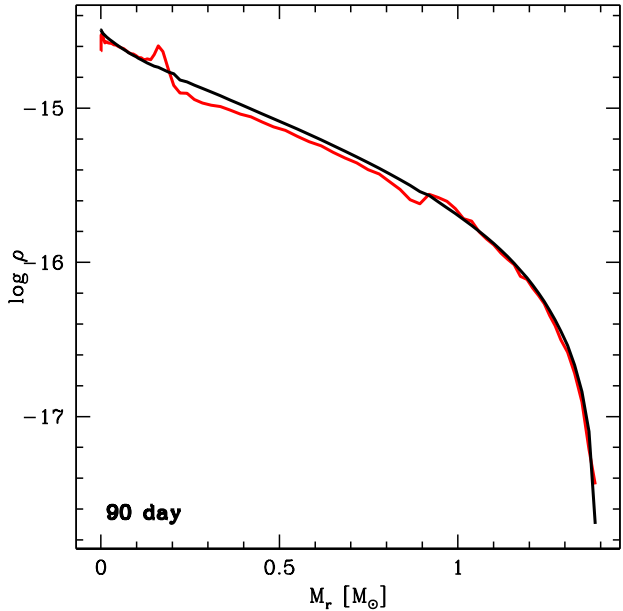}
\caption{Logarithm of the density profile of unmixed Model C070203
  computed by {\sc stella} at $t=90$ days (red lines) compared to the
  structure obtained by homologous expansion from the initial model
  (black lines).
\label{Fig:density_stella}}
\end{center}
\end{figure}

The photometry of most of the models, including all of them with the
standard ``M-mixing'', was also independently verified using an older
multi-energy group radiation hydro code {\sc stella}
(\citealt{Bli98,blsoruv}).  {\sc stella} was developed primarily for
Type II supernovae, where the effects of coupling of radiation
transfer to the hydrodynamics are more important, e.g., during shock
propagation \citep{Bli03,Chu04}.  {\sc stella} is not a Monte Carlo
code, but employs a direct numerical solution of radiative transfer
equation in the moment approximation.  Time-dependent equations for
the angular moments of intensity in fixed frequency bins are coupled
to the Lagrangian hydro-equations and solved implicitly.  The photon
energy distribution may be quite arbitrary.  Due to coupling with
hydrodynamics, which is generally not needed for the coasting phase of
SN~Ia (though see below), the radiative transfer part of the
calculation is somewhat cruder than in {\sc sedona}.

The effect of line opacity is treated by {\sc stella}, in the current
work, as an expansion opacity according to \citet{Eas93}, similar to
{\sc sedona}. The line list is limited to $\sim$160 thousand entries,
selected from the strongest down to the weakest lines until 
saturation in the expression for expansion flux opacity is achieved.
This is, in general, much less than in {\sc sedona}, moreover, the list
is optimized for solar abundance only.

The ionization and atomic level populations are described by
Saha-Boltzmann expressions.  However, the source function is {\em
not\/} in complete LTE.  
The source function at wavelength $\lambda$
is
\begin{equation}
S_\lambda = \epsilon_{\rm th}\,B_\lambda + (1-\epsilon_{\rm th})\,J_\lambda,
\label{Eq:S}
\end{equation}
where $\epsilon_{\rm th}$ is the thermalization parameter, $B_\lambda$
the Planck function, and $J_\lambda$ the angle mean intensity.  In
this work, {\sc stella} assumes $\epsilon_{\rm th}=1$.  Hydrodynamics
coupled to radiation is fully computed (homologous expansion is not
assumed).  This question, and major other recent improvements in the
code {\sc stella}
\citep[introduced after the paper][was published]{Bli98} are described
in (\citealt{Bli06}).

The heating by the decays of $^{56}$Ni $\to$ $^{56}$Co$\to$ $^{56}$Fe
is taken into account. It is assumed that positrons, born in the
decays, are trapped so they deposit their kinetic energy locally.  To
find the radioactive energy deposition, we treat the gamma-ray opacity
as a pure absorptive one, and solve the gamma-ray transfer equation in
a one-group approximation following~\cite{SSH}.

The effective opacity is assumed to be $\kappa_\gamma=0.05 Y_e \;
\mbox{cm}^2/\mbox{g}$, where $Y_e$ is the total electron number
density divided by the baryon density.  Although this approach is
checked against other algorithms, e.g. those used in {\sc eddington}
\citep{Bli98} it is less accurate than the full Monte-Carlo treatment in
{\sc sedona}

To calculate SNe~Ia light curves {\sc stella} can use up to 200
frequency bins and up to $\sim$400 zones in mass as a Lagrangian coordinate on
a modest processor, but all current results are obtained with 100 groups in
energy and 90 radial mesh zones.

As described by \citet{Bli98}, {\sc stella} has only an approximate
treatment of light travel time correction, since it works not with
time-dependent intensity, but with time-dependent energy and fluxes
only.  {\sc sedona} is superior in this aspect since it works directly
with packets of photons.

The approximation of homologous expansion is usually exploited in the
radiative transfer codes that neglect hydrodynamics
(\citealt{Eas93,Lucy05,Kasen_Sedona}).  However, \citet{Pin00a} point
out that the energy released in the \Nifs\ decay can influence the
dynamics of the expansion. The \Nifs\ decay energy is
$3\times10^{16}$ \ergg, which is equivalent to a speed of $2.5\times
10^3$ km s$^{-1}$, if transformed into the kinetic energy of a gram of
material. \citet{Pin00a} state: ``Since the observed expansion
velocity of SNe~Ia is in excess of $10^4$ \kms, we expect that this
additional source of energy will have a modest, but perhaps not
completely negligible, effect upon the velocity structure.''  In
reality, the heat released by the \Nifs\ decay will not all go into
the expansion of the SN~Ia.  If the majority of \Nifs\ is located in
the central regions of the ejecta then the main effect is an increase
in the entropy and local pressure (both quantities are dominated by
photons in the ejecta for the first several weeks).  The weak
overpressure will lead to a small decrease in density at the location
of the `nickel bubble', as well as to some acceleration of matter
outside the bubble.  It is often said that the expansion of the ejecta
is supersonic and that pressure cannot change the velocity of the
matter, but one should remember that in the vicinity of each material
point we have a `Hubble' flow, so differential velocities are in fact
{\em subsonic} in a finite volume around each point.

All of the models considered here have moderately mixed distributions
of $^{56}$Ni.  Other models (without mixing) demonstrate that the
nickel bubble, i.e. the depression of density in $^{56}$Ni-rich
layers, continues to grow during the coasting stage.  The effect is
modest, but as expected it may result in a $\sim$$10$\% difference in
velocity and density. This effect is evident, e.g., for Model C070203
(see Fig.~\ref{Fig:density_stella}).  The solid black line is the
initial model scaled to our result at 90 days since explosion. We see
that the density profile has changed due to the $^{56}$Ni and
\Cofs\ decays; it clearly deviates from homology.

The change in the density is important for the deposition of gamma-ray
energy, which is reflected in the light curve.  This may explain
partly some of the differences between {\sc stella} and {\sc sedona}
results.  The larger the \Nifs\ mass, the larger is the difference
expected.  


\section{CODE AND MODEL VERIFICATION AND VALIDATION}

\subsection{Comparison of Results from Sedona and Stella}
\lSect{codecomp}

\begin{figure}
\begin{center}
\includegraphics[clip=true,width=\columnwidth]{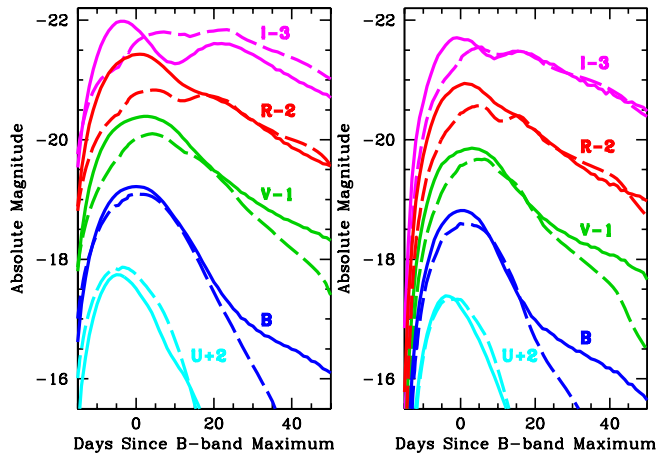}
\caption{Comparison of the $UBVRI$ light curves calculated by {\sc
    stella} (dashed line) and {\sc sedona} (solid line) for Models
    M070103 (left) and M040303 (right) Agreement for the $B$-band
    light curve during the first 20 days after maximum, which is of
    greatest interest in this paper, is excellent for the models that
    produce medium to large quantities of \Nifs\ and still quite good
    for those with low values like M040303. The divergence in the
    models after day 25 is due primarily to the lesser amount of lines
    used by {\sc stella} to compute the expansion opacity, and may
    also be influenced by the different treatment of the Ca~II IR
    triplet lines \citep{Kasen_IRLC}.
\label{Fig:compare}}
\end{center}
\end{figure}

Figure~\ref{Fig:compare} shows the comparison of the $UBVRI$ light
curves calculated by {\sc stella} and {\sc sedona} for Models M070103
and M040303. Agreement for the $B$-band light curve during the first
20 days after maximum, which is of greatest interest in this paper, is
quite good for the models that produce medium to large quantities of
\Nifs, and still reasonable for those with low values like
M040303.  In general, the results of the two codes are mutually
confirming, however one does note moderate differences in the $B$-band
light curve decline rates and peak magnitudes, and substantial
differences in the early $I$ and $R$-band light curves and the late
time $B$ and $V$-band light curves.  Note that the results of both
codes become increasingly unreliable after 40 days past $B$-band
maximum, due to the increasing inadequacy of the LTE
excitation/ionization assumption.

Our numerical experiments suggest that these differences in the
radiative transfer results are explained primarily by the different
atomic line data used in the separate codes.  As discussed in
\cite{Kasen_IRLC}, the use of an extensive atomic line list (with 
$\ga 5$~million lines) is critical in synthesizing the light curves of
SNe~Ia, especially in the red and near-infrared wavelength bands.
Because the number of weak lines treated in
\MCcode\ is much larger than that in {\sc stella},
the results of the former code show a generally superior
correspondence with observations.  However, even the more extensive
atomic line list used in \MCcode\ is likely still somewhat incomplete
and inaccurate.  Inadequacies in the available atomic data remain an
important source of uncertainty in supernova light curve modeling.

Other differences between the two transfer codes may also contribute
to the discrepancies seen in Figure~\ref{Fig:compare}.
\MCcode\ employs a finer frequency grid, and thus better resolves
individual spectral features.  The features can have an important
impact on the broadband magnitudes.  In addition, \MCcode\ treats the
Ca~II IR triplet lines as pure scattering, which likely better
represents the source function.  Because the Ca~II lines become
excessively strong at later epochs, they can affect the radiative
transfer in all bands \citep{Kasen_IRLC}.  \MCcode\ also includes a
more detailed, multi-group gamma-ray transfer procedure, and thus more
accurately determines the radioactive energy deposition rate and late
times.  The main advantage of {\sc stella} is its self-consistent
treatment of hydrodynamics, but arguments in the previous section and
the density plot Fig.~~\ref{Fig:density_stella} suggests that
deviations from homology are not large.  For present purposes,
\MCcode\ is superior and most of the subsequent figures and discussion
in this paper are based on the results of that code, except where
otherwise noted.

\subsection{Comparison With Analytic Approximations and Other Calculations}
\lSect{W7}

\begin{figure}
\begin{center}
\includegraphics[clip=true,width=\columnwidth]{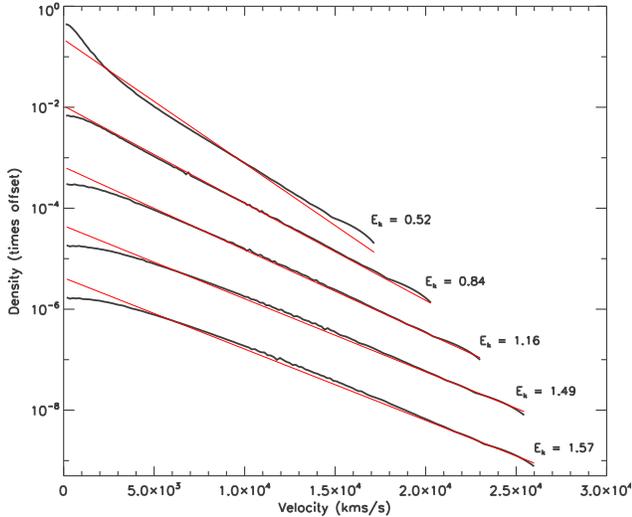}
\caption{Logarithm of the density structure of explosion models with
  different kinetic energy (black lines) compared to the exponential
  function (Equation~\ref{Eq:exp_density}, red lines).  The kinetic
  energy \KE\ of each model is marked on the figure.
\label{Fig:density}}
\end{center}
\end{figure}

\begin{figure}
\begin{center}
\includegraphics[clip=true,width=\columnwidth]{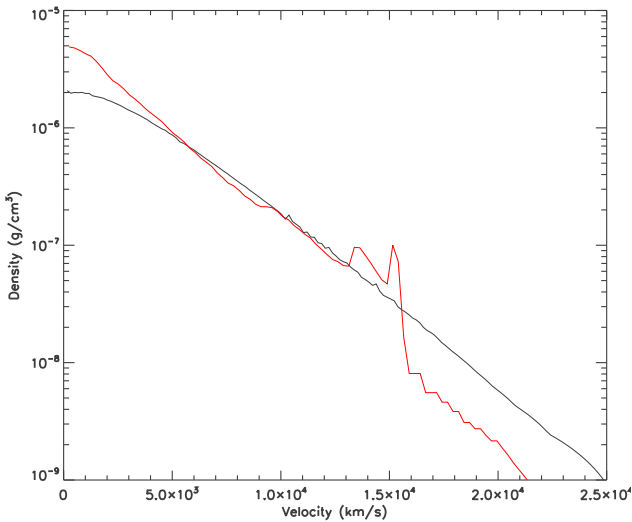}
\caption{Logarithm of the density structure at 10000~sec for Model
  M060303 (black line) compared to that of W7 (red line).
\label{Fig:w7_density}}
\end{center}
\end{figure}

The primary output of the hydrodynamic explosion calculation is the
final density structure (e.g., density versus velocity) of the ejecta
once it has reached the homologous phase.  Interestingly, the final
density profiles of all models are well characterized by a simple
exponential function which depends only upon the kinetic energy of
the explosion
\begin{eqnarray}
\rho(v,t) = \frac{\Mch}{8\pi v_e^3 t^3} \exp[-v/v_e]  
\\\mathrm{where}~~v_e = \sqrt{\frac{E_k}{6 \Mch}} = 
2455 \biggl(\frac{E_k}{10^{51}}\biggr)^{1/2}~\kms 
\label{Eq:exp_density}
\end{eqnarray}
Figure~\ref{Fig:density} shows that this simple analytic formula holds
very well for all but the innermost ejecta ($v \la v_e$).  For the
high \KE\ models, the formula overestimates the central densities,
while from the low \KE\ models it underestimates them.

The properties of our models can be compared to existing standard
SN~Ia explosion models. A much studied 1-D model that has been shown
to agree with typical observed light curves and spectra is Model W7 of
\citet{Nom84} \citep[see also][]{Thielemann_w7,Iwa99}. In terms of our four
composition parameters, the abundances in the final frozen-out Model
W7 are $^{56}$Ni, 0.59 \msun, stable iron ($^{54,56}$Fe, $^{55}$Mn,
$^{58,60}$Ni), 0.26 \msun, and Si-Ca, 0.27 \msun. If our simple model
for the hydrodynamics is correct and if mixing is not a major issue,
this should be a good match for our Model M060303 (which has 0.6, 0.3,
and 0.3 \msun \ of $^{56}$Ni, stable iron and IME respectively).

In Fig.~\ref{Fig:w7_density}, we compare the density profile of
Model M060303 to that of W7.  The agreement is quite good in the inner
layers, although there are discrepancies for velocities $v >
13,000$~\kms.  This is the region in which the deflagration burning
began to be quenched in W7, which lead to the production of a density
spike. This density spike would probably be absent in a
multi-dimensional simulation.

\begin{figure}
\begin{center}
\includegraphics[clip=true,width=\columnwidth]{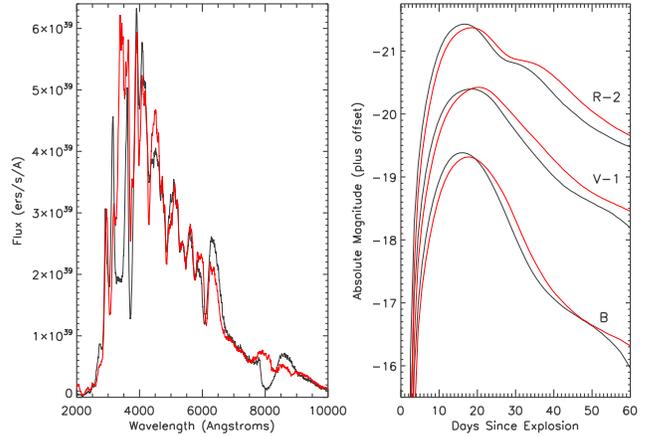}
\caption{Demonstration that the simplified Model M060303 (black lines)
  well reproduces the observable properties of the well-known W7 model
  (red lines).  \emph{Left:} Comparison of the near maximum light
  ($\texp = 18$~day) spectra of the two models.  \emph{Right:}
  Comparison of the $BVR$-band light curves of the two models.
\label{Fig:w7_compare}}
\end{center}
\end{figure}

In Fig.~\ref{Fig:w7_compare} we compare the light curve and near
maximum light spectra (\texp = 18~days) of Model M060303 and W7.
Despite the differences in the density profile and approximate
representation of the composition in M060303, the overall agreement is
reasonably good. The light curves of M060303 are slightly faster than
W7, due to the slightly lower densities and greater stable iron group
production (see \S\ref{sec:stablefe}).  The maximum light spectra show
only minor discrepancies in the lines of calcium and in the
ultraviolet reflecting the different mixing employed here and the fact
that the Ca abundance in Model M060303 is about twice as large as in
W7 (0.024 \msun {\sl vs} 0.012 \msun) and extends to higher velocity.

\subsection{Comparison to Individual Observations}
\lSect{compobs}

\begin{figure}
\begin{center}
\includegraphics[clip=true,width=\columnwidth]{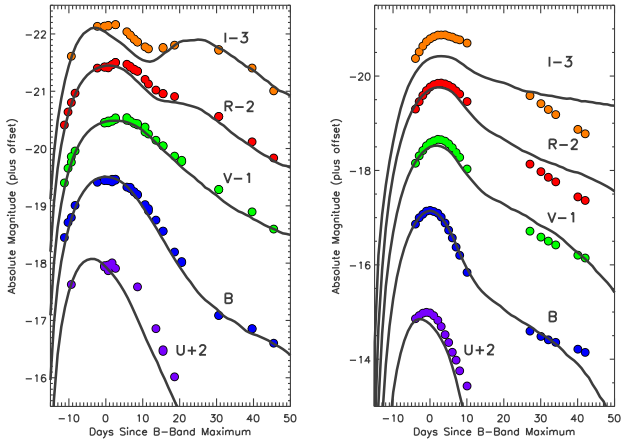}
\caption{\emph{Right:} Comparison of the broadband synthetic light
  curves of Model M080202 to observations of the bright/broad Type~Ia
  SN~1991T \citep{Lira_91T}.  \emph{Left:} Comparison of the broadband
  synthetic light curves of Model M010309 to observations of the
  Type~Ia SN~1999by, a subluminous SN~1991bg-like event
  \citep{Garnavich_99by}.
\label{Fig:LC_peculiar}}
\end{center}
\end{figure}

\begin{figure}
\begin{center}
\includegraphics[clip=true,width=\columnwidth]{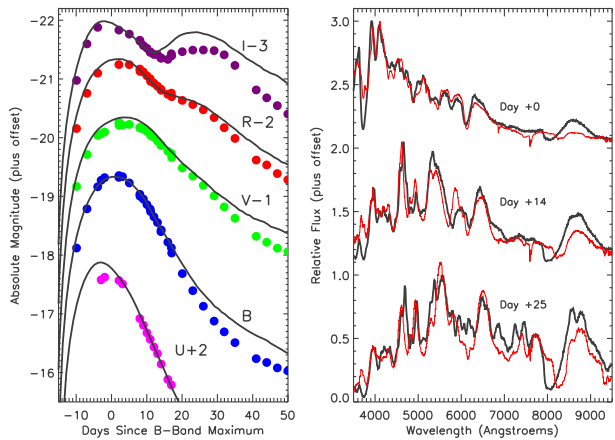}
\caption{Verification of the Model M070103 against SN~Ia observations.
  \emph{Left:} Synthetic broadband light curves of the model (solid
  lines) compared to observations of the normal Type~Ia SN~2001el
  \citep[][filled circles]{Kris_01el}.  In order to improve the visual
  comparison, the observations have been offset by $-0.1$~mag in order
  to better align them with the model light curves.  \emph{Right:}
  Synthetic spectra at three different epochs with respect to $B$-band
  maximum of the model (black lines) compared to observations of the
  normal Type~Ia SN~1994D (red lines).
\label{Fig:obs_compare}}
\end{center}
\end{figure}

\begin{figure}
\begin{center}
\includegraphics[clip=true,width=\columnwidth]{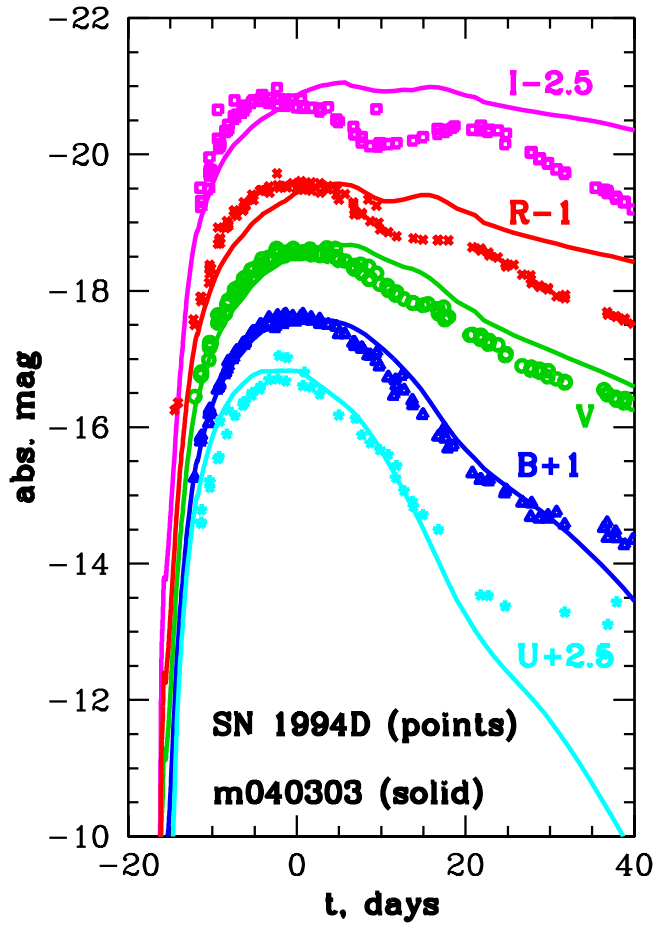}
\caption{Comparison of the broadband synthetic light curves of Model
  M040303, computed by {\sc stella} (solid lines), to observations of
  the Type~Ia SN~1994D \citep{richmond95,meikle96}
\label{Fig:LC_stella}}
\end{center}
\end{figure}

Within our model grid are examples that resemble a broad range of
observed SNe~Ia, from photometrically normal events like SN~2001el, to
bright, broad events like SN~1991T, and faint, narrow ones like
SN~1991bg.  In Fig.~\ref{Fig:obs_compare}, the $UBVRI$ light curves
Model M070103 ($\Mni = 0.7~\msun$) are compared with those of the
normal Type~Ia SN~2001el \citep{Kris_01el}.  The observed light curves
of SN~2001el have been corrected for dust extinction using the
estimates $A_v = 0.57$ and $R_v = 2.88$, and the distance modulus is
taken to be $\mu = 31.3$ \citep{Kris_01el, Wang_01el}.  Generally good
agreement between model and observations is found in all bands.  The
model further reproduces the double-peaked behavior of the I-band
light curve, although the secondary maximum is stronger than in the
observations.

Figure~\ref{Fig:LC_peculiar} demonstrates that our models also include
examples resembling more extreme SNe~Ia events.  The left panel of the
figure compares the light curves of Model M080101 ($\Mni = 0.8~\msun$) to those
of the overly bright SN~1991T \citep{Lira_91T}.  We adopt the
reddening estimate of $A_v = 0.50.~R_v = 3.1$ from
\cite{Phillips_1999} and the Cepheid distance measurement of $\mu =
30.56$ from \cite{Gibson_91T}, although there is substantial
uncertainty in these values.  The right panel of the figure compares
the light curves of Model M010207 ($\Mni = 0.1~\msun$) to those of the
subluminous SN~1991bg-like event SN~1999by.  The reddening and
distance here are taken to be $A_v = 0.43,~R_v = 3.1,~\mu = 30.75$
\citep{Garnavich_99by}.  Good agreement with observations is found for both 
examples.

Full time-series of synthetic spectrum calculations are available from
the \MCcode\ code and Fig.~\ref{Fig:obs_compare} (right panel)
compares the Model M070103 spectra at several epochs to observations
of the normal Type~Ia SNe~1994D \citep{meikle96,Patat_94D}.  While
there are some differences in detail, the model reproduces quite well
the essential spectroscopic features and colors over a wide evolution
time.  The only major discrepancy occurs in the Ca~II IR-triplet
features, which is too strong in the model.  This confirms the
adequacy of the transfer calculations to model the spectral and color
evolution of SNe~Ia over the timescale of interest.

Fig.~\ref{Fig:LC_stella} compares synthetic light curves of Model
M040303, computed by {\sc stella}, to observations of the Type~Ia SN~1994D
\citep{richmond95,meikle96}. The agreement is especially good in $UBV$, and in general it is not worse than for the MPA deflagration models presented by \citet{Bli06}.
Comparison with {\sc sedona} results for the same model in
Fig.~\ref{Fig:compare} suggests that the agreement with observations
in $R$ and $I$ filters can be improved by extending the line list used in
{\sc stella} for computing expansion opacity.

\section{BASIC PHYSICS OF THE WIDTH-LUMINOSITY RELATION}
\lSect{basicwlr}

Before turning to the model calculations, it is useful to summarize
the basic radiative transfer physics relating to the brightness and
decline rate of SN~Ia light curves.  A more detailed discussion of the
transfer effects can be found in a companion paper \citep{Kas06}.
Because the light curves of SNe~Ia are influenced by a number of
physical parameters, the behavior of our model light curves can not
usually be explained by referencing a single cause, but rather require
consideration of a combination of interrelated transfer effects.  We
attempt to describe the most important effects individually below.

It is well known that the light curves of SNe~Ia are powered entirely
by the decay of radioactive \Nifs\ (and its daughter $^{56}$Co)
synthesized in the explosion.  The mass of \Nifs\ produced (\Mni) is
therefore the primary determinate of the peak brightness of the event.
On the basis of approximate analytic models, \citep{Arn82} showed
that the bolometric luminosity at peak is roughly equal to the
instantaneous rate of radioactive energy deposition
\begin{equation}
L_p \sim f \Mni \exp(-t_p/t_{Ni})
\label{Eq:Arnett}
\end{equation}
where $t_p$ is the rise time to peak, $t_{Ni} \approx 8.8$~days is the
\Nifs\ decay time and $f$ is the percentage of the gamma-ray
decay energy that is trapped at the bolometric peak (typically $f
\ga 0.9$).  

In SNe~Ia, the ejecta remain optically thick for the first several
months after explosion.  The width of the bolometric light curve is
related to the photon diffusion time.  The basic diffusion physics can
roughly be understood using simple scaling arguments. In a standard
random walk, the diffusion time is given by $t_d \sim R^2/\lambda_p c$
where $R$ is the radius, $\lambda_p = 1/\kappa \rho$ is the photon
mean free path, and $\kappa$ is the mean opacity.  In homologous
expansion $R = \phiNi v t$ where $v$ is the characteristic ejecta
velocity, $t$ is the time since explosion, and
\phiNi\ is a factor describing the fractional distance between the
bulk of \Nifs\ and the ejecta surface, roughly: $\phiNi \sim (M -
\NiCent)/M$ where \NiCent\ is the center of mass of the \Nifs\
distribution.  Using the scaling relations for the total ejected mass
$M \sim \rho v^3 t^3$ and kinetic energy $\KE \sim M v^2$, one finds
\begin{equation}
t_d \sim \phiNi\ \kappa^{1/2}  M^{3/4} \KE^{-1/4},
\label{Eq:lc_width}
\end{equation}
The important parameters affecting the bolometric diffusion time are
thus the total mass $M$, the kinetic energy \KE, the radial
distribution of nickel \phiNi, and the effective opacity per unit
gram, $\kappa$.  The last of these is the most complicated, depending
upon the composition, density, and thermal state of the ejecta, as
well as the velocity shear across the ejecta.  In general, $\kappa$
increases with temperature/ionization, thus models with larger \Mni\
will typically have slightly longer diffusion times
\citep{Khokhlov_93, Pin01, Hoeflich_99by, Kas06}

A further factor influencing the bolometric light curve is the rate at
which the ejecta become transparent to gamma-rays from radioactive
decay.  Near maximum light, the densities in a Chandrasekhar-mass
model are high enough that nearly all gamma-rays are trapped locally
($f \ga 0.9$).  However, by 15 days after maximum densities have
dropped such that a substantial percentage of the gamma-rays escape
the ejecta without being thermalized.  Models in which the bulk of
\Nifs\ is located further out in mass coordinates will experience a
more rapid transition to gamma-ray transparency and hence possess a
generally faster bolometric decline rate.

In addition to the parameters affecting the bolometric decline rate
just mentioned, one must also consider the physics affecting the
spectroscopic and color evolution of SNe~Ia.  In \citet{Kas06}, it was
shown that the WLR arises primarily from a
\emph{broadband} effect.  In particular, the $B$-band light curve decline rate
depends sensitively on the rate at which the SN colors shift
progressively redwards following maximum light.  Dimmer SNe~Ia
(i.e., those with lower \Mni) exhibit a generally faster color
evolution, which is the primary reason for their faster $B$-band
decline.  Physically, this reflects the faster
\emph{ionization evolution} of dimmer SNe~Ia.  Following
maximum-light, the SN colors are increasingly determined by the
development of numerous Fe~II and Co~II lines that blanket the bluer
wavelength bands and, at the same time, increase the emissivity at
longer wavelengths.  Because dimmer SNe~Ia are generally cooler, they
experience an earlier onset of Fe~III to Fe~II recombination in the
iron-group rich layers of ejecta.  Consequently, Fe~II and Co~II line
blanketing develops more rapidly in dimmer SNe~Ia, resulting in a more
rapid evolution of the SN spectral energy distribution to the red.
This is the principle explanation for their faster $B$-band decline
rate.

As a corollary, one realizes that the velocity distribution of iron
group elements plays an additional important role in determining the
broadband light curves.  Models in which iron group elements are
concentrated at low velocities are unable to form strong Fe~II/Co~II
line features in the post-maximum epochs, and consequently will
exhibit a slower color evolution (and hence $B$-band decline rate)
compared to those in which the iron group elements are mixed out to
higher velocity.

%
\section{RESULTS}
\lSect{results}

\subsection{All Models Combined}

\begin{figure}
\begin{center}
\includegraphics[clip=true,width=\columnwidth]{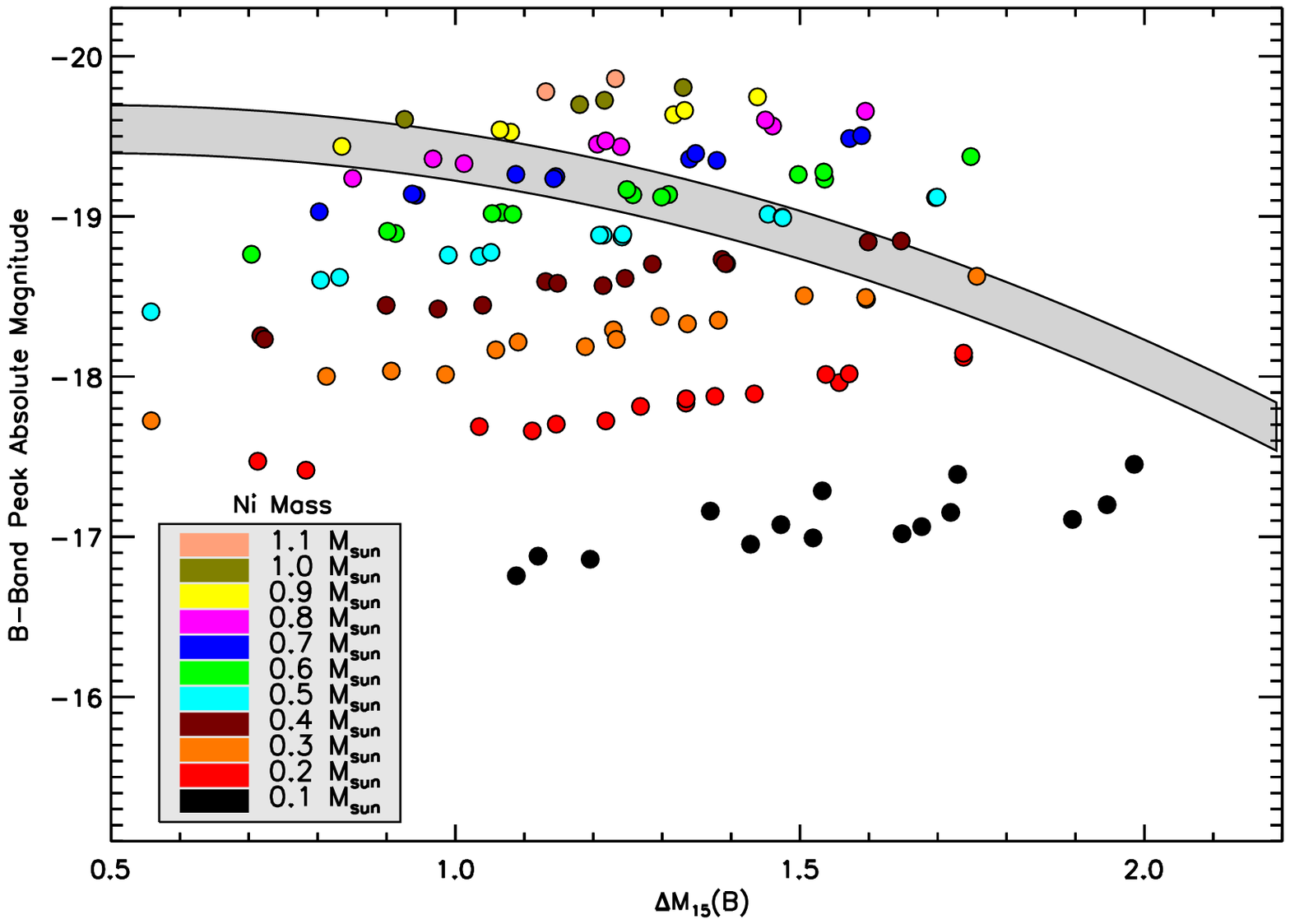}
\caption{Relationship between the $B$-band decline rate \dmfb\ and
  peak $B$-band magnitude for the full set of M-series models. The
  color coding shows the \Nifs\ mass, which varies from $0.1$ to
  $1.1$~\msun.  The shaded region is the observed width-luminosity
  relation of \cite{Phillips_1999} with a calibration $\Mb = -19.3$
  for $\dmfb = 1.1$ and a dispersion of $\sigma = 0.15$~mag.  In
  contrast to the observations, the models occupy a wide region in the
  plot, indicating a sensitivity of the light curves to parameters
  other than \Nifs.The systematic offset of $\sim0.06$ magnitudes in color
  between the observations and models suggests that assumed
  thermalization parameter $\epsilon_{\rm th} = 1$, likely
  overestimates the actual redistribution probability.
\label{Fig:all_ni}}
\end{center}
\end{figure}

\begin{figure}
\begin{center}
\includegraphics[clip=true,width=\columnwidth]{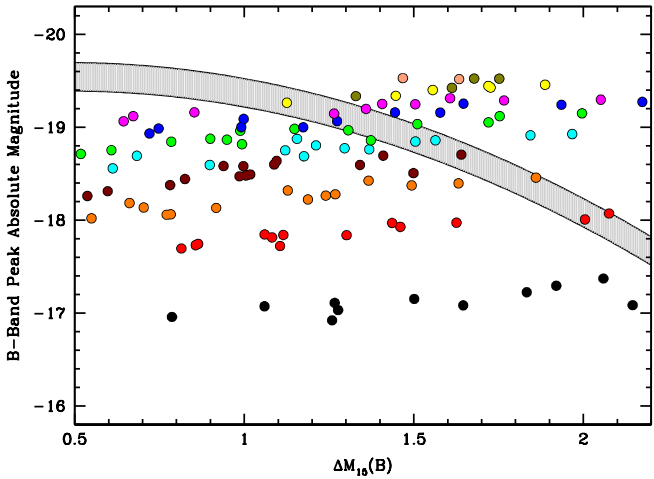}
\caption{Relationship between the $B$-band decline rate \dmfb\ and
  peak $B$-band magnitude for the full set of M-series models computed
  by {\sc stella}.  The color coding is the same as in
  Fig.\protect\ref{Fig:all_ni}.
\label{Fig:all_ni_stella}}
\end{center}
\end{figure}

\begin{figure}
\begin{center}
\includegraphics[clip=true,width=\columnwidth]{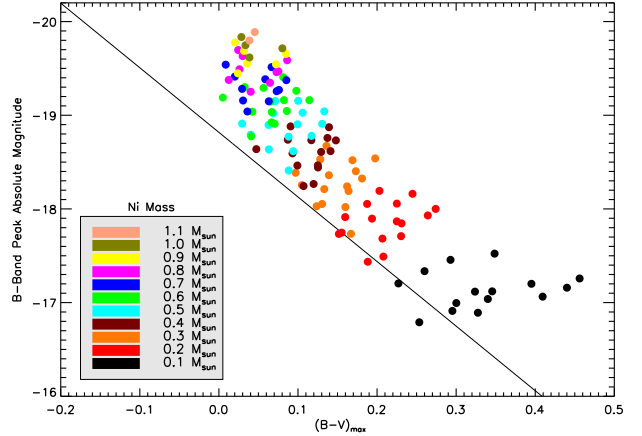}
\caption{Relationship between the maximum light $B$-$V$ color and the
  peak $B$-band magnitude for the full set of M-series models. The
  color coding shows the \Nifs\ mass, which varies from $0.1$ to
  $1.1$~\msun.  The solid line is the observed relation of
  \cite{Phillips_1999}, with a calibration $\Mb = -19.3$ for $\dmfb =
  1.1$.
\label{Fig:all_bmv}}
\end{center}
\end{figure}

The WLR is often quantified as relation between peak $B$-band magnitude
\Mb\ and the drop in $B$-band magnitude 15 days after peak \dmfb.
Figures~\ref{Fig:all_ni} and \ref{Fig:all_ni_stella} show the WLR for
the full set of moderately mixed models calculated with the {\sc
sedona} and {\sc stella} codes, respectively.  The models are
color-coded by the mass of
\Nifs\ each produces as defined in Fig.~\ref{Fig:all_ni}. Overplotted
in both figures, as a shaded band, is the empirical WLR of
\cite{Phillips_1999}, for which an absolute calibration of $\Mb =
-19.3$~mag at $\dmfb = 1.1$~mag and a dispersion of $\sigma =
0.15$~mag \citep{Ham96} have been adopted.

In contrast to the observed behavior, the models span a wide region in
the figure.  {\sl Contrary to popular expectations, the light curve
width and luminosity are not both determined by a single parameter,
the $^{56}$Ni produced in the explosion}. If this were the case, all
models with a given mass of $^{56}$Ni would collapse to one point on
the plot, and that point would be in the gray band.

As the plots confirm, \Mni\ is clearly the dominant parameter
affecting the SN peak magnitude.  For a given \Mni\ value, \Mb\ varies
only by about $\pm 0.25$~mag.  The decline rate \dmfb, on the other
hand, spans at least a full magnitude for a given \Mni\ mass,
indicating its sensitivity to additional physical parameters.  Given
the scaling of the diffusion time (Equation~\ref{Eq:lc_width}), one
can anticipate that two very important parameters are the total
kinetic energy of the explosion, \KE, and the radial distribution of
\Nifs, \phiNi.  This is confirmed in the following sections by
examining suitable subsets of models.

Careful examination of the figures shows, in fact, that for fixed
\Mni, the model light curves actually exhibit an ``anti-Phillips
relation'', i.e., the brighter supernovae are more narrow.  This is
not a surprising result, as SNe with shorter diffusion times (i.e.,
faster light curves) lose a smaller percentage of their internal
energy to adiabatic expansion, and thus reach a brighter peak
earlier. Equivalently, this can understood as an expression of
Arnett's rule (Equation~\ref{Eq:Arnett}).  Models with broader light
curves typically have a longer rise time $t_p$ and thus, for given
\Mni, are dimmer at peak.

Although the models in Figs.~\ref{Fig:all_ni} and
\ref{Fig:all_ni_stella} do not reproduce the observed WLR relation,
they nonetheless lead to a very important physical insight -- the true
SN~Ia explosion mechanism realizes only a small subset of the
theoretically conceivable possibilities, implying a rather tight
internal correlation between the relevant physical parameters.  This
places a strong constraint on theoretical explosion paradigms.
Indeed, in the following we use this constraint to deduce some of the
properties of the SN~Ia ejecta structure.

The large spread seen in Figs.~\ref{Fig:all_ni} and
\ref{Fig:all_ni_stella} also suggests that intrinsic SN variation is
likely a significant source of intrinsic scatter in the WLR.  Because
\dmfb\ depends upon other parameters than \Mni, any uncorrelated
variation of the secondary parameters leads to dispersion in the
WLR. Any systematic variation of the parameters with progenitor
environment could form a potential basis for evolutionary effects.

Interestingly, if one plots the peak model magnitudes of the entire
set {\sl vs} the $B$-$V$ at maximum instead of \dmfb, a tighter
correlation results (Fig.~\ref{Fig:all_bmv}).  The slope of the model
correlation closely resembles that of the observed relation given in
\cite{Phillips_1999}. The models are systematically redder than the
observations by $\sim 0.06$~mag, suggesting that our assumed
thermalization parameter ($\epsilon_{\rm th} = 1$) likely
overestimates the true redistribution probability in SNe~Ia 
(\Sect{danscode}).  Though it is clear that not all the models shown
in Fig.~\ref{Fig:all_bmv} are frequently realized in nature, the
smaller dispersion suggests that such color indicators may be less
sensitive to intrinsic variation in the supernovae.

\subsection{Sources of Dispersion}

The large spread in the model WLR of Figs.~\ref{Fig:all_ni} and
\ref{Fig:all_ni_stella} indicates that parameters other than \Mni\
significantly affect the light curves of SNe~Ia.  We show below that
the most important of these are the total burned mass, \Mburn, which
determines the \KE, and the stable iron mass, \Mfe, which influences
\phiNi.  In addition, the degree of direct \Nifs\ mixing is significant
as well.  For a given \Mni\, these parameters have significant impact
on the decline rate \dmfb, and hence may act as sources of dispersion
in the WLR.  Here the effects of each are quantified using suitable
subsets of the models

\subsubsection{Effect of the Total Burned Mass}
\lSect{constm}

\begin{figure}
\begin{center}
\includegraphics[clip=true,width=\columnwidth]{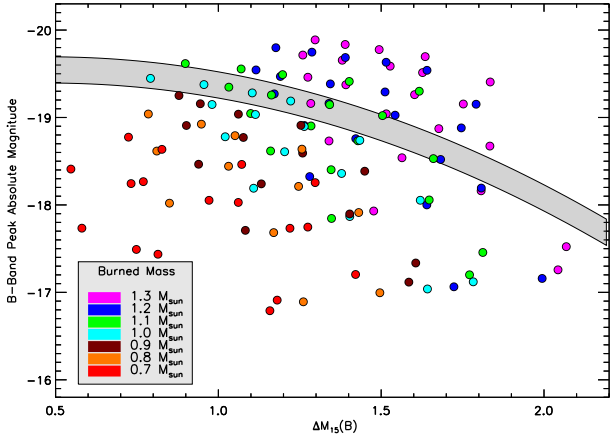}
\caption{Same as Fig.~\ref{Fig:all_ni}, but color coded to show the
  total burned mass of the models. Models with the same total
  explosion energy, i.e., points of the same color {\sl do} roughly
  yield a width-luminosity relation in which ``brighter equals
  broader''. The best agreement is for a burned
  mass of $1.1 \pm 0.1$ \Msun, though the narrower light curves are
  somewhat fainter than the observations if burned mass is a
  constant.
\label{Fig:all_mburn}}
\end{center}
\end{figure}

\begin{figure}
 \begin{center}
\includegraphics[clip=true,width=\columnwidth]{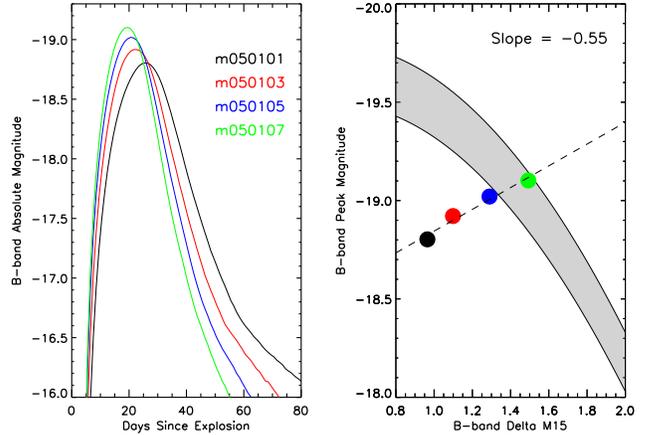}
\caption{Effect of the total burned mass (and hence kinetic energy) on
  the $B$-band light curve for a subset of M-series models with
  constant \Nifs. The models shown have \Mni = 0.5~\msun\ and \Mfe =
  0.1~\msun\ and \Mime = $[0.1,0.3,0.5,0.7~\msun]$.  The left panels
  shows the $B$-band model light curves, while the right panel shows
  the corresponding width-luminosity relation (as in
  Fig.~\ref{Fig:all_ni}).  For a constant \Nifs\ mass, models with a
  greater total burned mass have higher kinetic energy, and hence
  faster light curves. All else being equal, the full width of the
  observed region corresponds to a range in burned mass of only 0.2
  \Msun.
\label{Fig:mbu_models}}
\end{center}
\end{figure}

The models shown in Figs.~\ref{Fig:all_ni} and \ref{Fig:all_ni_stella}
span a wide range in the fraction of the original white dwarf
which is burned to heavier elements.  Because the nuclear energies
released in burning carbon and oxygen to iron, $^{56}$Ni or IME are
all about the same, and since the same initial white dwarf is used for
all calculations, the total mass burned in the explosion, \Mburn,
essentially dictates the final kinetic energy, \KE, of the ejecta.
Given the inverse dependence of the diffusion time on \KE\
(Equation~\ref{Eq:lc_width}), a higher value of \Mburn\ can be
expected to lead to a relatively faster bolometric light curve for any
given model.  In addition, because the velocity of the bulk of \Nifs\
ejected in the explosion increases with
\KE, a higher \Mburn\ should also lead to a faster evolution of the
model colors and gamma-ray transparency, both of which further
contribute to a faster $B$-band light curve decline (\Sect{basicwlr}).

Figure~\ref{Fig:all_mburn} redisplays the same full model set as in
Figs.~\ref{Fig:all_ni} and \ref{Fig:all_ni_stella}, but now color
coded by the total burned mass.  Models with higher \Mburn\ (high \KE)
generally occupy the faster declining portion of the plot while models
with lower \Mburn\ have overly broad light curves.  It is clear from
the figure that a large part of the spread in the model WLR is due to
the variations in \Mburn\ among the models.  Eliminating both the very
high and very low \Mburn\ points does result in a loose inverse
correlation between peak brightness and decline rate (i.e., brighter
is general broader), though there is still a large dispersion compared
with observations. {\sl The best agreement with observations is
achieved if common SN Ia burn 1.0 to 1.2 \msun \ of their mass to
silicon and heavier elements.} Without further modification, models
that burn 0.7, 0.8, and 1.3 \msun\ seem to be rare events.

We can further quantify the effect of \Mburn\ on the light curves by
examining a subset of models with fixed $\Mni = 0.5$~\msun\ and $\Mfe
= 0.1~\msun$, but with \Mime\ varied from 0.1 to 0.7~\msun.  The total
burned mass among these models thus varies from \Mburn\ =
0.7-1.3~\msun\ corresponding to a variation of \KE\ from 0.68 to
1.43~B.  Figure~\ref{Fig:mbu_models} shows that the models obey an
anti-Phillips relation.  Models with higher \Mburn\ have shorter rise
times and faster declines, and are also brighter at peak.
Quantitatively, increasing the amount of burned IME mass by 0.2~\msun\
(a \KE\ increase of 0.25~B) increases \Mb\ by 0.1~mag, \dmfb\ by
0.2~mag, and decreases the $B$-band rise time by 3~days.

\subsubsection{Effect of the Mass of stable iron and Mixing}
\lSect{stablefe}

\begin{figure}
\begin{center}
\includegraphics[clip=true,width=\columnwidth]{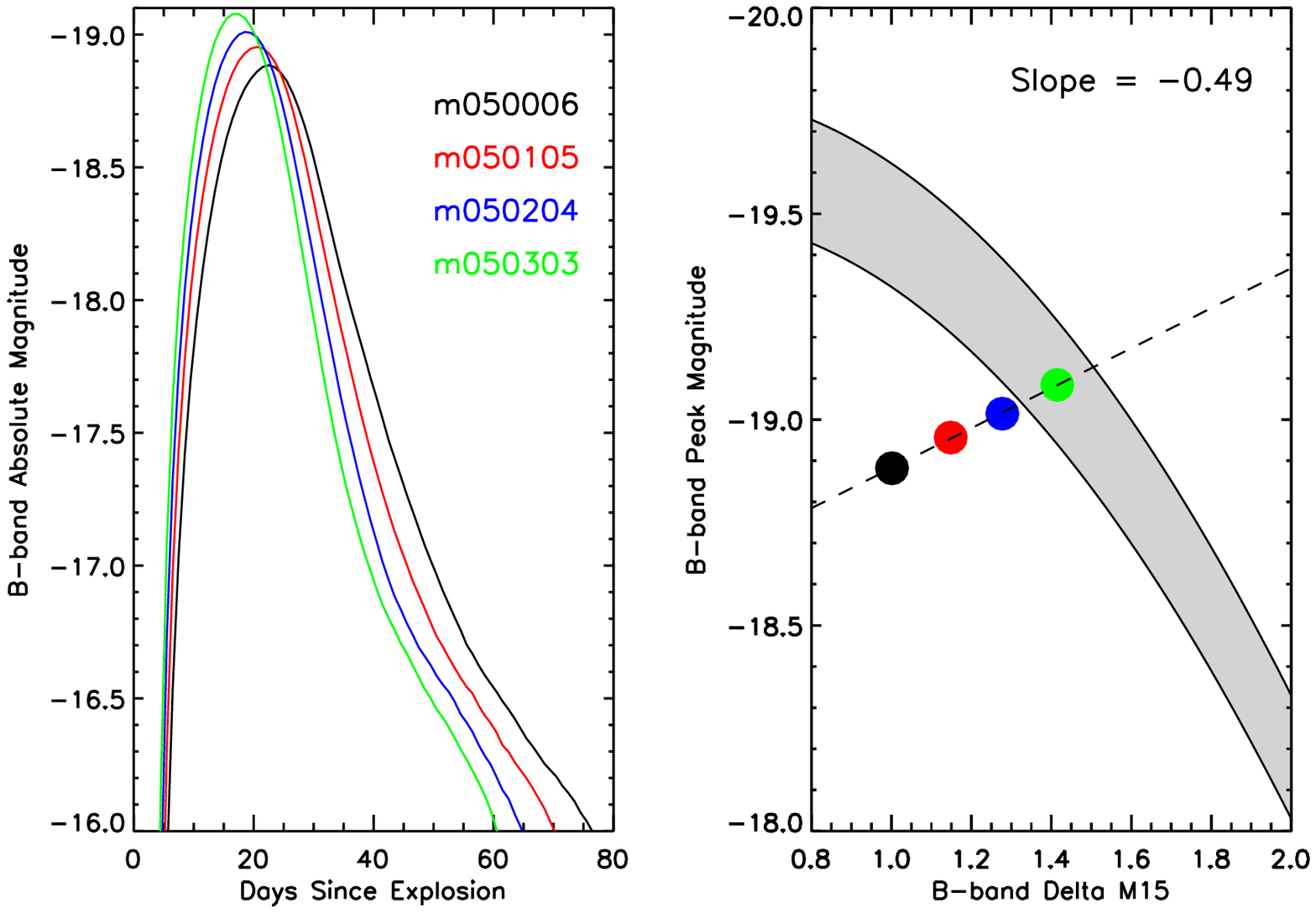}
\caption{Effect of changing the mass of stable iron-group elements
  (\Mfe) on the $B$-band light curve for a subset of M-series
  models. Same as Fig.~\ref{Fig:mbu_models}, expect that here the mass
  of \Nifs\ and the explosion energy are held constant ($\Mburn =
  1.1~\Msun$, in all cases), while the amount of stable iron is varied
  at the expense of IME. Here \Mni = 0.5~\msun, \Mfe =
  $[0.0,0.1,0.2,0.3~\msun]$ and $\Mime + \Mfe = 0.6~\msun$.  Models
  with larger \Mfe\ have \Nifs\ distributed closer to the surface, and
  hence faster light curves. A change in \Mfe\ of only 0.1
  \Msun\ gives the entire width of the observed band.
\label{Fig:mfe_models}}
\end{center}
\end{figure}

\begin{figure}
\begin{center}
\includegraphics[clip=true,width=\columnwidth]{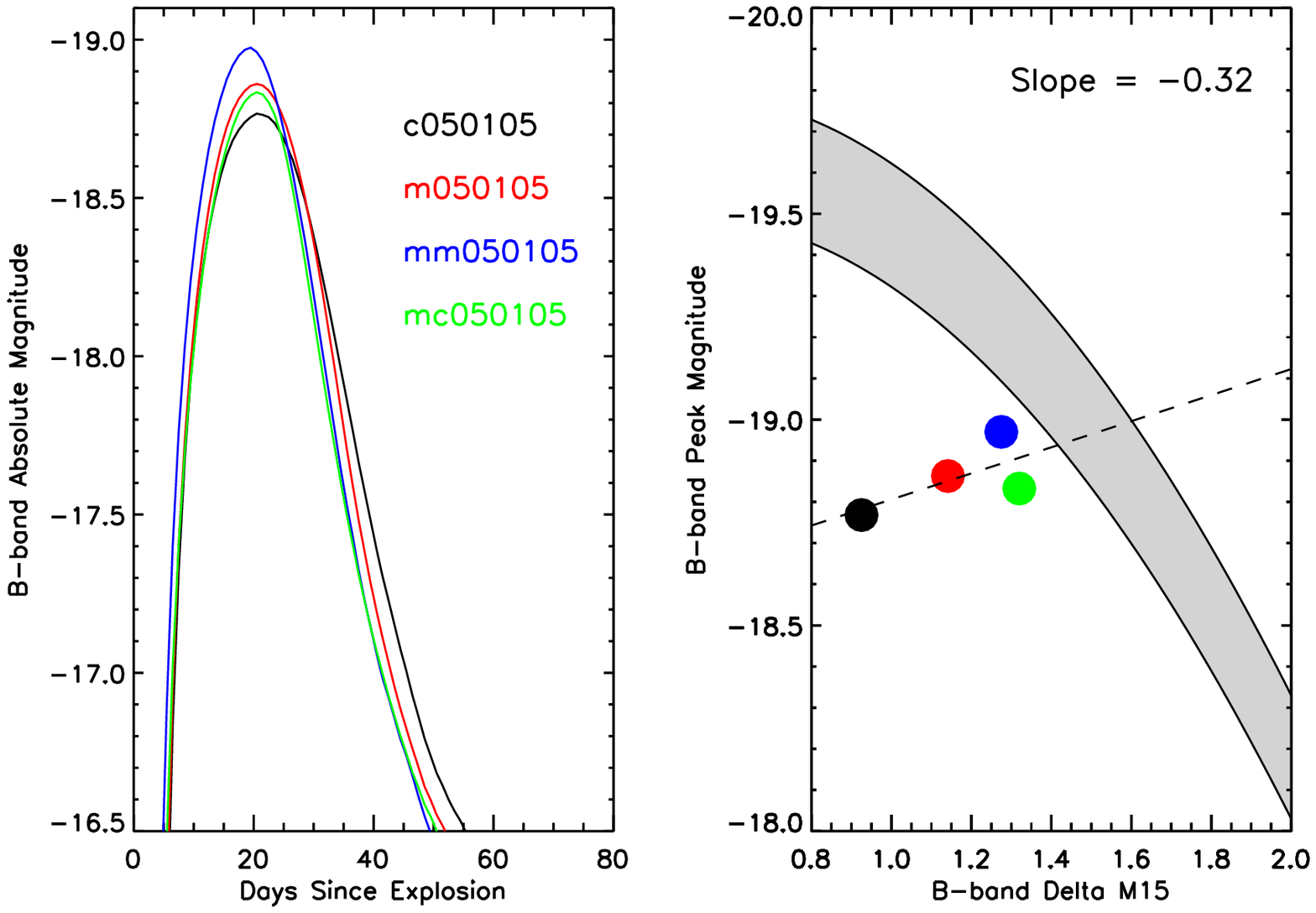}
\caption{Effect of the degree of ``mixing'' on the $B$-band light
  curve.  Same as Fig.~\ref{Fig:mbu_models}, but for four models all
  with the same ejected composition, $\Mni = 0.5~\msun$, $\Mfe =
  0.1~\msun$, $\Mime = 0.5~\msun$, but each with different mixing
  prescriptions.  The more heavily mixed models have
  \Nifs\ distributed closer to the surface and hence faster light
  curves.
\label{Fig:mix_models}}
\end{center}
\end{figure}

The second important model parameter leading to the dispersion in
Figs.~\ref{Fig:all_ni} and \ref{Fig:all_ni_stella} is the mass of
stable iron, \Mfe. Since it is not a source of radioactive decay
energy, and since its opacity is not so different from \Nifs, the
chief effect of \Mfe\ is to influence the distribution of \Nifs.  For
models with larger values of {\sl central} \Mfe, the \Nifs\ center of
mass is pushed farther out, thus decreasing the \phiNi\ parameter
(this is not necessarily the case for $^{54}$Fe and $^{58}$Ni produced
{\sl in} the \Nifs\ zones because of a finite metallicity and neutron
excess).  This can be expected to lead to faster light curves for
three reasons discussed in \Sect{basicwlr}: (1) based upon
Equation~\ref{Eq:lc_width}, the diffusion time to the ejecta surface
should be shorter; (2) the occurrence of \Nifs\ at lower densities
leads to a lower percentage of gamma-ray trapping in the post-maximum
epochs; and (3) the increase in iron group elements at higher velocity
layers of ejecta leads to the stronger development of Fe~II/Co~II
features in the post-maximum spectra, contributing to a faster color
evolution (and hence $B$-band decline rate).

To demonstrate the important effect of \Mfe\ in
Fig.~\ref{Fig:mfe_models}, models were selected with fixed \Mni =
0.5~\msun\ and fixed \Mburn = 1.1~\msun, but with \Mfe\ varied from
0.0 to 0.3~\msun.  The models also obey an anti-Phillips relation -- for
given \Mni, models with larger \Mfe\ have faster light curves.
Quantitatively, increasing \Mfe\ by 0.1~\msun\ increases \Mb\ by
0.05~mag, \dmfb\ by 0.1~mag, and decreases the $B$-band rise time by
2~days.

These effects of \Mfe\ on the light curves are not related to the
presence of stable iron group elements \emph{per se}, but to the
effect \Mfe\ has on the distribution of \Nifs.  Essentially the same
effect can be demonstrated more directly by varying the degree of
mixing in the model.  In Fig.~\ref{Fig:mix_models} we show four
models each with the same compositional production ($\Mni =
0.5~\msun$, $\Mfe = 0.1~\msun$, $\Mime = 0.5~\msun$) but each with
different degrees of mixing.  The more heavily mixed models have a
greater proportion of \Nifs\ in the outer layers of ejecta, and hence
faster light curves, for the same three reasons given above.

\section{TOWARDS A WORKING MODEL}
\lSect{cuts}

Having identified the two principal physical parameters that cause
dispersion in the model WLR -- the explosion energy and the
distribution of \Nifs\ in the ejecta, selected subsets of the models are
now examined that are in better accordance with the observations.
What are the common properties of those ``viable'' models that fall
within the observed WLR in Figs.~\ref{Fig:all_ni} and
\ref{Fig:all_ni_stella}, and why is the observed scatter so small?

\subsection{Constraints from Nucleosynthesis and Nuclear Physics}
\lSect{nuconstrain}

\begin{figure}
\begin{center}
\includegraphics[clip=true,width=\columnwidth]{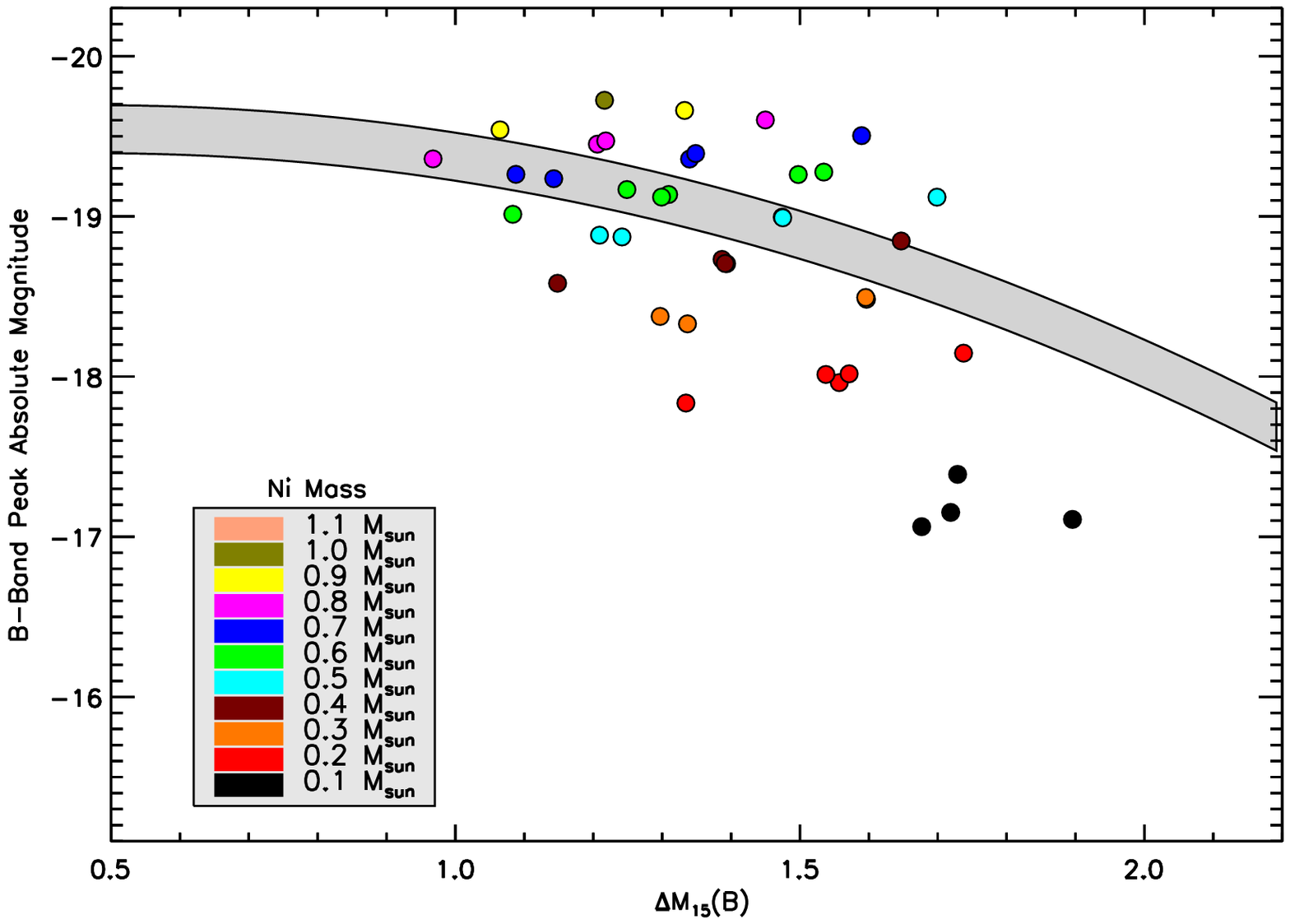}
\caption{Same as Fig.~\ref{Fig:all_ni}, but including only those
  models which obey the observational constraints of rapid expansion
  and ``reasonable'' nucleosynthesis and spectra, $\Mburn = 1.1 \pm
  0.1$~\msun, $\Mime \ge 0.1~\msun$, $\Mfe = 0.1-0.3~\msun$.
\label{Fig:all_constrain}}
\end{center}
\end{figure}

\begin{figure}
\begin{center}
\includegraphics[clip=true,width=\columnwidth]{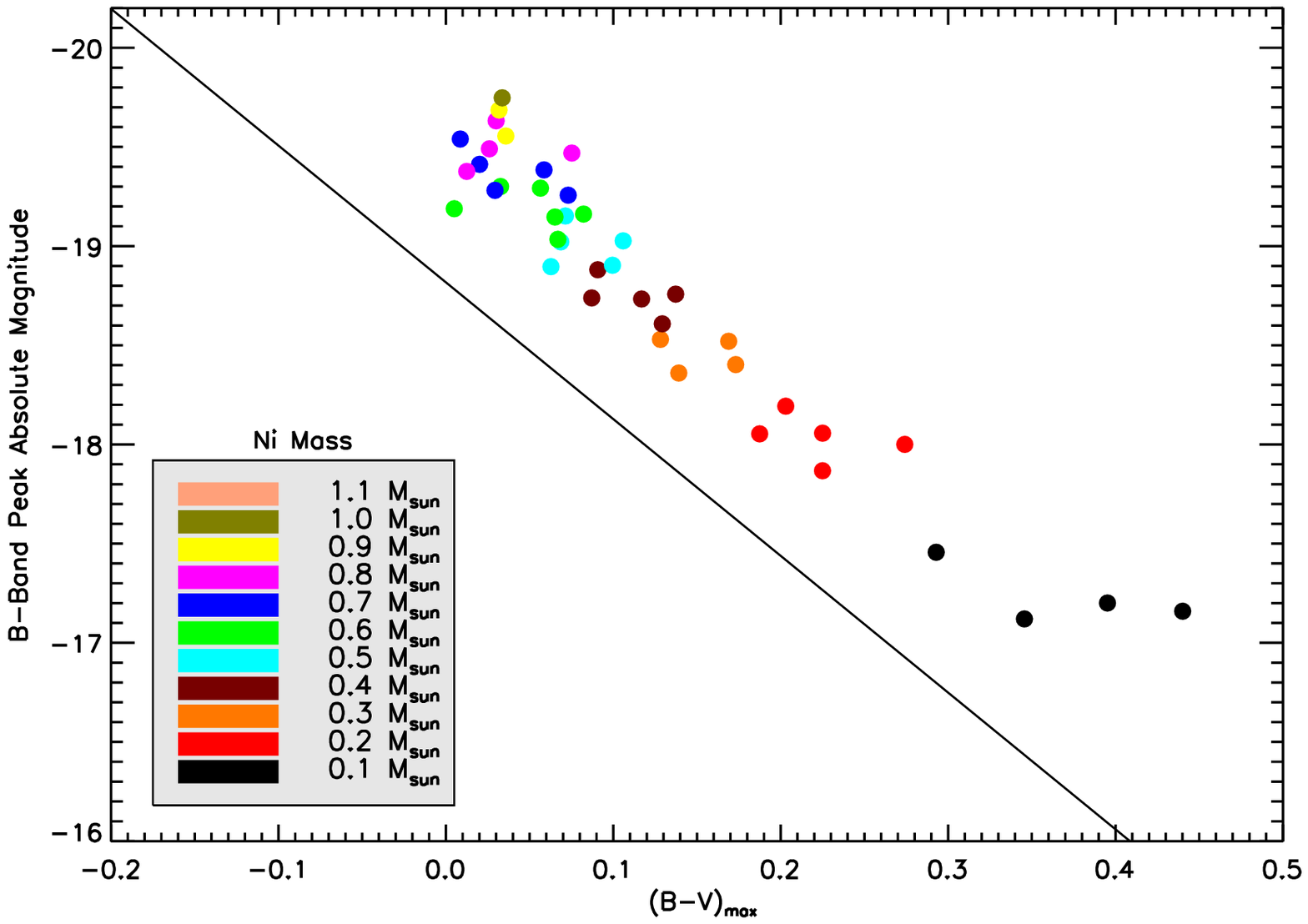}
\caption{Same as Fig.~\ref{Fig:all_bmv}, but including only those
  models which obey the observational constraints of rapid expansion
  and ``reasonable'' nucleosynthesis and spectra, $\Mburn = 1.1 \pm
  0.1$~\msun, $\Mime \ge 0.1~\msun$, $\Mfe = 0.1 - 0.3~\msun$.  Once
  again, the solid line is the observed relation of
  \cite{Phillips_1999}, with a calibration $\Mb = -19.3$ for $\dmfb =
  1.1$. The systemmatic offset of $\sim0.06$ magnitudes in color
  between the observations and models suggests that assumed
  thermalization parameter $\epsilon_{\rm th} = 1$, likely
  overestimates the actual redistribution probability.
\label{Fig:bmv_constrain}}
\end{center}
\end{figure}

First, one might consider what is ``reasonable'' from other quarters.
Nature does not just select, with equal frequency, any random
combination of nuclear products that unbind the star, but must select
values consistent with known nuclear physics and, on the average, with
the requirements of stellar nucleosynthesis and SN Ia spectroscopy
(e.g., the presence of IME in the spectrum).  Does the application of
these observational constraints serve to increase the agreement of
the models with the WLR?

It is not reasonable, for example, that a SN Ia make no stable iron,
but only \Nifs. Ignition in models near the Chandrasekhar mass can
only be achieved at densities in excess of about $2.5 \times 10^9$ g
cm$^{-3}$ and any burning near that density makes $^{54,56}$Fe and
$^{58,60}$Ni, not \Nifs. Further, if the star has any appreciable
metallicity, the excess neutrons will mostly end up in these same
stable iron-group isotopes. As \citet{Tim03} discuss, all initial CNO
will end up at the end of helium burning in the isotope $^{22}$Ne,
creating a mass fraction approximately 0.02 (Z/\zsun). Subsequent
burning and conservation of neutrons turns this chiefly into $^{54}$Fe
and $^{58}$Ni with a combined mass fraction $\approx 0.05$ Z/\zsun.
For a range of \Nifs\ masses up to 1 \msun, and metallicities Z = 0 to 3
\zsun, this implies a stable iron mass of 0 to 0.15 \msun. An
additional {\sl minimum} of 0.1 \msun \ of $^{54}$Fe is expected from
electron capture. This is roughly the amount of burning required to
reduce the white dwarf density below the point where electron capture
is important; \citep[e.g.,][]{Nom84}. Thus we expect \Mfe $\gtaprx$
0.1 \msun\ always.

On the other hand, nucleosynthesis {\sl in the Milky Way Galaxy}
requires that the sum of $^{54}$Fe and $^{58}$Ni be approximately 10\%
by mass that of $^{56}$Fe. Given that massive stars also make some
iron, this might possibly be raised to $\sim$15\%, but the iron in
Type II supernovae is made in a region of high neutron excess as
well. This suggests that Galactic SN Ia do not, on the average, make
more than 0.2 \msun \ of stable iron per event.  Of course, one does
not know the isotopic composition of iron in distant galaxies.  It is
reasonable, however, that whatever physical constraints operate to
limit the amount of electron capture in local SN Ia also function in
similar explosions far away.  In summary, it seems that the stable
iron mass is restricted by nuclear physics and nucleosynthesis to
typically 0.1 to 0.3 \msun\ even for metallicities as high as three
times solar.  This does not mean there cannot occasionally be
supernovae with very different characteristics. Nucleosynthesis
constraints only operate on the average.

The total mass burned is also limited by observational constraints on
the typical expansion speed. The supernovae do not come apart with
only a small excess of total energy over the binding energy, or
spectral lines would be too narrow and ionization stages too neutral.
In terms of light curves, it is already known from
Fig.~\ref{Fig:all_mburn} that approximately 1.1 \Msun \ needs to
burn. This means that the typical model burns most of its mass. That
is natural in all detonation models, and also true of strong
deflagrations.

Finally, the presence of strong silicon, sulfur and calcium lines in
the maximum-light spectrum of SNe~Ia implies the production of at
least 0.1~\msun\, and probably 0.2~\msun\ of IME for normal events.
For these elements, there is no nucleosynthetic upper bound and the
spectroscopic limits are not presently highly constraining.

Figure~\ref{Fig:all_constrain} and Figure~\ref{Fig:bmv_constrain} show
the resulting plots of \Mb {\sl vs} \dmfb\ and \Mb\ {\sl vs} $B$-$V$
when these constraints are applied.  In particular, we choose those
models with $\Mburn = 1.1 \pm 0.1$~\msun, $0.1 \ge \Mfe \le 0.3$~\msun
and $\Mime \ge 0.1$~\msun.  Unlike Fig.~\ref{Fig:all_ni}, the more
restricted models do show a WLR, albeit a noisy one. The
scatter in the color-brightness plot is also reduced. Nature
apparently realizes this limited set of models more frequently.  The
WLR exists, not because a single parameter, the
\Nifs\ abundance determines both the brightness and decline rate of SN
Ia, but because of other physics that constrains the production and
distribution of stable iron, \Nifs, and IME.

The correlation between color and peak magnitude continues to be
tight in Fig.~\ref{Fig:bmv_constrain}, though the models are offset to
the red about 0.06 magnitudes. This again probably reflects a
thermalization efficiency of less than 100\% (\Sect{danscode}).

\subsection{A Constant Mass of Iron Group Elements}
\lSect{constfe}

\begin{figure}
\begin{center}
\includegraphics[clip=true,width=\columnwidth]{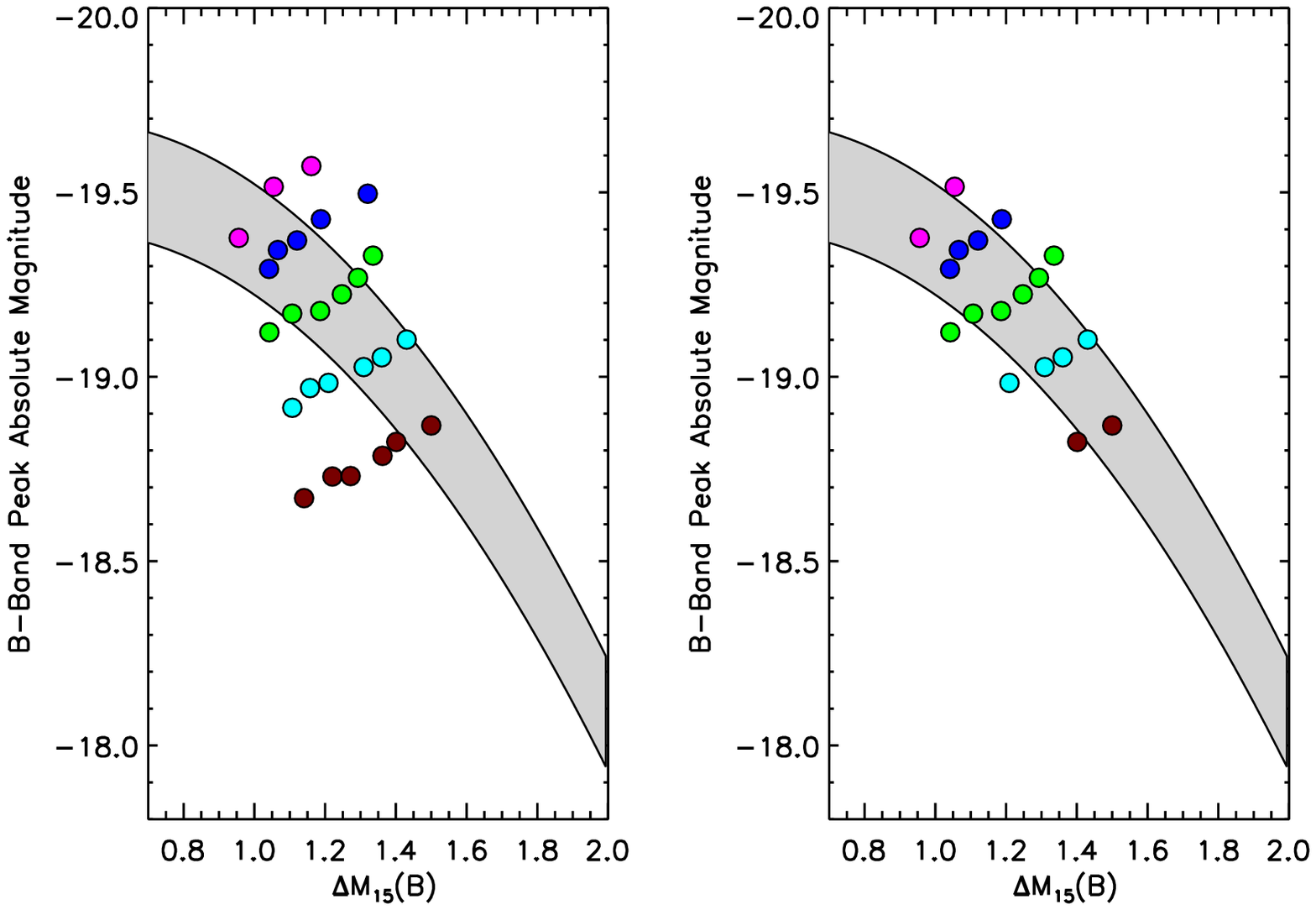}
\caption{\emph{Left:} WLR for a more finely gridded set of the mildly
  mixed (M-series) of models which obey the constraints $\Mburn =
  1.0-1.1~\msun$, $\Mfe = 0.1-0.3$~\Msun, and $\Mime \ge 0.1$~\Msun.
  The \Nifs\ mass varies from $\Mni = 0.4-0.8$~\Msun, with the same
  color coding as in Figure~\ref{Fig:all_ni}.  \emph{Right:} Same as
  left, but including only those models that have total iron group
  production in the range $\Mfe + \Mni = 0.7-0.9~\msun$.
\label{Fig:m_finer}}
\end{center}
\end{figure}

\begin{figure}
\begin{center}
\includegraphics[clip=true,width=\columnwidth]{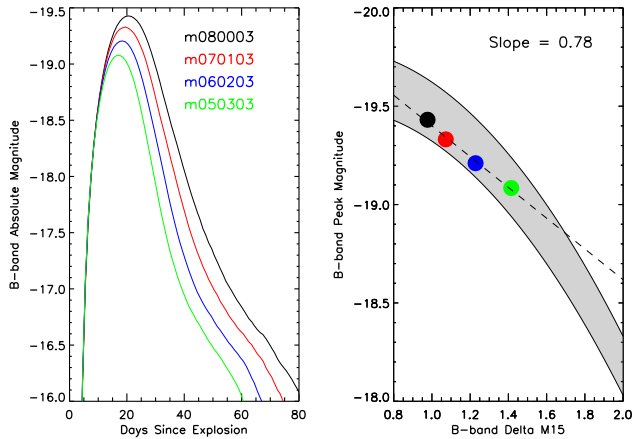}
\caption{Effect of varying \Mni\ while keeping the total iron group
  production ($\Mni + \Mfe$) fixed. The models shown have $\Mni =
  [0.5,0.6,0.7,0.8]$~\msun, $\Mime = 0.3$~\msun, and $\Mfe + \Mni =
  0.8$~\msun.  Such a variation might be attributed to a variable
  amount of electron capture and metallicity in an otherwise standard
  supernova.  However, Model M080003, which has no neutronized iron,
  is not realistic.  The total range of metallicity and
  electron-capture effects is probably bounded by the red and green
  points, i.e., about 0.3 magnitudes. The range of metallicity effects
  alone is less.
\label{Fig:fecore_models}}
\end{center}
\end{figure}

\begin{figure}
\begin{center}
\includegraphics[clip=true,width=\columnwidth]{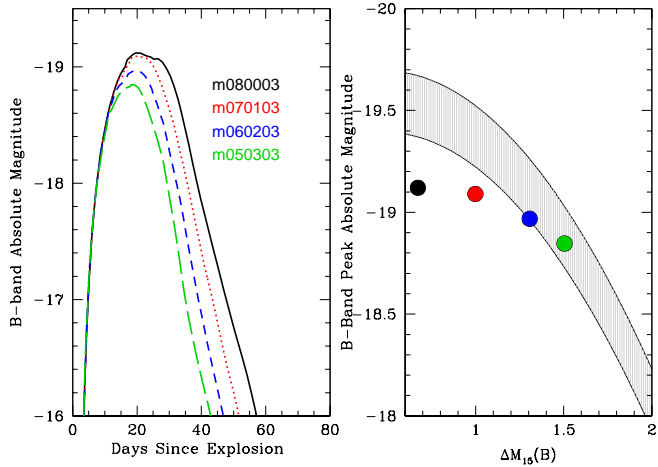}
\caption{ Same as Fig.~\ref{Fig:fecore_models}, but computed with {\sc
    stella}. A WLR of the correct sign is also present, though the
  agreement with observations for these particular models is not as
  good. For models with a small quantity of initial iron, especially
  M080003 and M070103, the light curve at early times is fainter in
  {\sc stella} because of a deficiency of lines included for Ni and Co
  (the Fe line list is more nearly complete). See text.
\label{Fig:fecore_stella}}
\end{center}
\end{figure}

\begin{figure}
\begin{center}
\includegraphics[clip=true,width=\columnwidth]{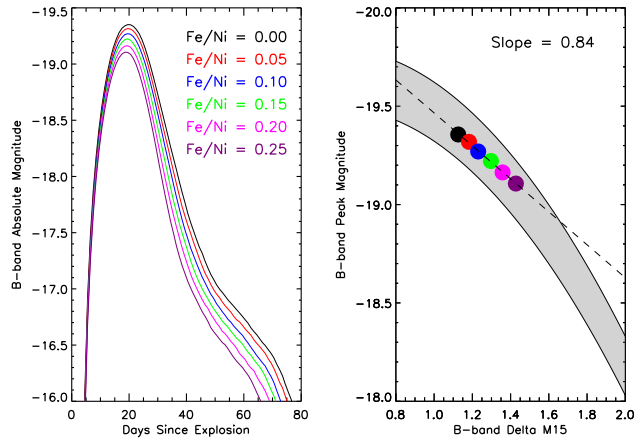}
\caption{Effect of varying the abundance of stable iron group
  throughout the \Nifs\ zone, as might be attributable to variations
  in progenitor metallicity.  The figure shows six models derived from
  Model M070103, in which the ratio of stable iron to \Nifs\ is varied
  throughout the \Nifs\ zone from 0 to 25\%.  All models have, in
  addition, 0.1~\Msun\ of stable iron at the center of the ejecta.
  The total mass of stable iron thus varies from 0.1 to 0.275~\Msun.
  The models follow the observed WLR.
\label{Fig:met_models}}
\end{center}
\end{figure}

The scatter of model points in Fig.~\ref{Fig:all_constrain} is still
much greater than observed.  To reduce it further, more stringent
restrictions must be applied.  In order to further populate the
allowed band, addition M-series models were constructed obeying the
following constraints: $\Mburn = 1.0 - 1.1~\msun$, $\Mfe =
0.1-0.3$~\Msun\ and $\Mime \ge 0.1$~\Msun.  A smaller range of \Nifs\
masses were also explored $\Mni = 0.4-0.8$~\Msun.
Figure~\ref{Fig:m_finer} (left panel) shows the width-luminosity
relation for this set of models.  The scatter is reduced compared to
the full plot, but is still larger than the observed.

A second cut of these models is then made assuming that the total mass
of iron group elements, as well as the total burned mass, must also
lie within a restricted range.  Figure~\ref{Fig:m_finer} (right panel)
shows only those models in which $\Mfe + \Mni = 0.7-0.9~\msun$, which
are seen to be the models that fall within the observed WLR. 

The necessity of having iron group elements in the region
0.7-0.9~\msun\ follows from the spectroscopic and color evolution effect
discussed in \cite{Kas06}.  The $B$-band decline rate is largely
determined by the rate at which Fe~II/Co~II line blanketing develops
in the post-maximum spectra.  In order for this line blanketing to
develop, there must be a high abundance of iron group elements out to
layers $v
\approx 8000~\kms$ which corresponds to mass coordinate $m =
0.8~\msun$.  This can also be achieved by mixing.

It is not unreasonable that the explosion produce a nearly constant
mass of iron group elements. The isotopic composition may vary because
of ignition density and metallicity, but all matter that burns with a
temperature over $\sim 5 \times 10^9$ K will be iron-peak isotopes of
some variety.  Apparently, not only do common SN Ia burn a nearly
constant fraction of their total mass, but a nearly constant fraction
of that achieves nuclear statistical equilibrium. Delayed detonation
at a nearly constant density (after a nearly constant amount of mass
has been burned) might be one way, but not the only way of achieving
that requirement.

\subsection{Variable Electron Capture and Metallicity}

The results of the last section suggest that SNe~Ia models in which
both the total iron group production and the total mass burned are
constants of the explosion are in better accord with the observed
WLR.  In such a scenario, the amount of $\Mni$ is varied at the
expense of stable iron group elements such as \Feff\ and \Nife.  There
are two ways in which this may come about.

First, stable iron group elements are produced at the center of the
ejecta from electron capture (embodied in our \Mfe\ parameter).  If
the central density of the white dwarf is higher, electron capture will
be enhanced and the ratio of \Mfe\ to \Mni\ will increase. Such
variations in ignition density might reflect variations in accretion
rate in binaries with different separation, masses, metallicity, etc.
A second mechanism that will vary the ratio \Mni\ to \Mfe\ is the
progenitor metallicity, which determines the relative abundance of
\Nifs\ to \Feff\ when burning to nuclear statistical equilibrium \citep{Tim03}. 

These two effects are difficult to separate in the model, though there
are observational constraints on the effect of metallicity
\citep{Gal05}. In the models, electron capture produces a more
centrally concentrated distribution of stable iron, while increased
metallicity makes the iron in the same place as the \Nifs. But the
effect on the light curve of increased electron capture plus extensive
mixing may be indistinguishable from that of increased metallicity
\citep[though see][mixing all the way to the surface has
  observational diagnostics. We have in mind mixing that does not
  extend to the outer layers.]{Hoe98}.

These two effects are demonstrated in Figs.~\ref{Fig:fecore_models},
\ref{Fig:fecore_stella}, and
\ref{Fig:met_models}. Figure~\ref{Fig:fecore_models} shows M-series
models in which \Mni\ is varied from $0.5$ to $0.8$~\Msun, but $\Mfe +
\Mni = 0.8~\msun$, and $\Mime = 0.1~\msun$ are both held constant.
These models, in which the stable iron remains concentrated at the
ejecta center (as expected from electron capture) are in reasonably
good agreement with the observed WLR.

Even better agreement is found if the stable iron is mixed throughout
the \Nifs\ zone, as expected from variations in metallicity.
Figure~\ref{Fig:met_models} shows a set of models derived from M070103
in which a constant ratio of \Feff\ to \Nifs\ is varied from 0 to 25\%
{\sl throughout} the \Nifs\ zone.  All models have, in addition,
0.1~\Msun\ of stable iron located at the ejecta center.  The
\emph{total} stable iron mass in these models thus varies from 0.1 to
0.275~\Msun.  The models fall very nicely along the observed WLR.

Thus variations in either (or, more likely, both) the progenitor
metallicity and the degree of electron capture can be a significant
source of the observed luminosity variations in SNe~Ia.  However, it
appears that this effect has a limited range. Varying the
metallicity from 0 to 3 times solar only leads to variations of 15\%
in \Mni. Larger values of metallicity may be unreasonable, and a
floor, $\sim0.1$ \Msun, on the lowest $^{54}$Fe abundance is set by
electron capture \citep[e.g.,][]{Nom84}. Less than 0.3 magnitudes of
the observed peak magnitudes, and perhaps 0.4 magnitudes in decline
rate can be explained this way \citep[see also][]{Hoe98}. This is
roughly one-third of the observed spread in peak magnitude for common
SN Ia, and is consistent with the observational limits of
\citet{Gal05} that metallicity is {\sl not} the major cause of the
width-luminosity relation, or of the preponderance of bright SN Ia in
late galaxies \citep{Ham96}. This is especially true, given that part
of the spread in Figs.~ \ref{Fig:fecore_models} and
\ref{Fig:met_models} is due to electron capture during the explosion.

Figure~\ref{Fig:fecore_stella} presents the same series of models as
Fig.~\ref{Fig:fecore_models}, but computed with {\sc stella}.
Reasonably good agreement persists for models that keep both the total
iron group production ($\Mni + \Mfe$) and IME production fixed, though
the agreement is less good for brighter, broader supernovae because
these models have no or little Fe in the beginning.  Although {\sc
  stella} has $\sim 10^5$ lines of Fe in the opacity, the line list is
probably too poor for Ni and Co.  Lower opacity means lower emission
according to Kirchhoff's law (which is applicable here when the line
opacity is treated as an absorptive one).  This makes the brightest
models underluminous.

We note that our results in this section differ from the conclusions
of \cite{Mazzali_06}, who also studied the effect of varying the ratio
of stable iron group species to \Nifs\ in SN~Ia models.  We find that
varying this ratio leads to models in general accord with the WLR,
whereas \cite{Mazzali_06} find that the variation creates dispersion
from it.  They therefore identify the ratio as a possible ``second
parameter'' in the WLR.  The different conclusions likely stem from
the different approaches to the radiative transfer problem.
\cite{Mazzali_06} use monochromatic LC calculations and a simplified
form of the mean opacity which depends only on the iron group
abundance and time.  These calculations therefore do not directly
capture the critical dependence of the model spectral/color evolution
on the ejecta temperature and ionization state. \cite{Mazzali_06} do
attempt to quantify the color effects using static synthetic spectra
at select epochs.  These spectral calculations, however, may be
limited by the adoption of an extended inner boundary surface which
emits blackbody radiation.  In this case, the continuum formation is
treated only approximately, and the radius of the inner boundary
surface becomes an important free parameter of the calculation.

\subsection{Constant $\Mni + \Mime$}
\lSect{nivsime}

\begin{figure}
\begin{center}
\includegraphics[clip=true,width=\columnwidth]{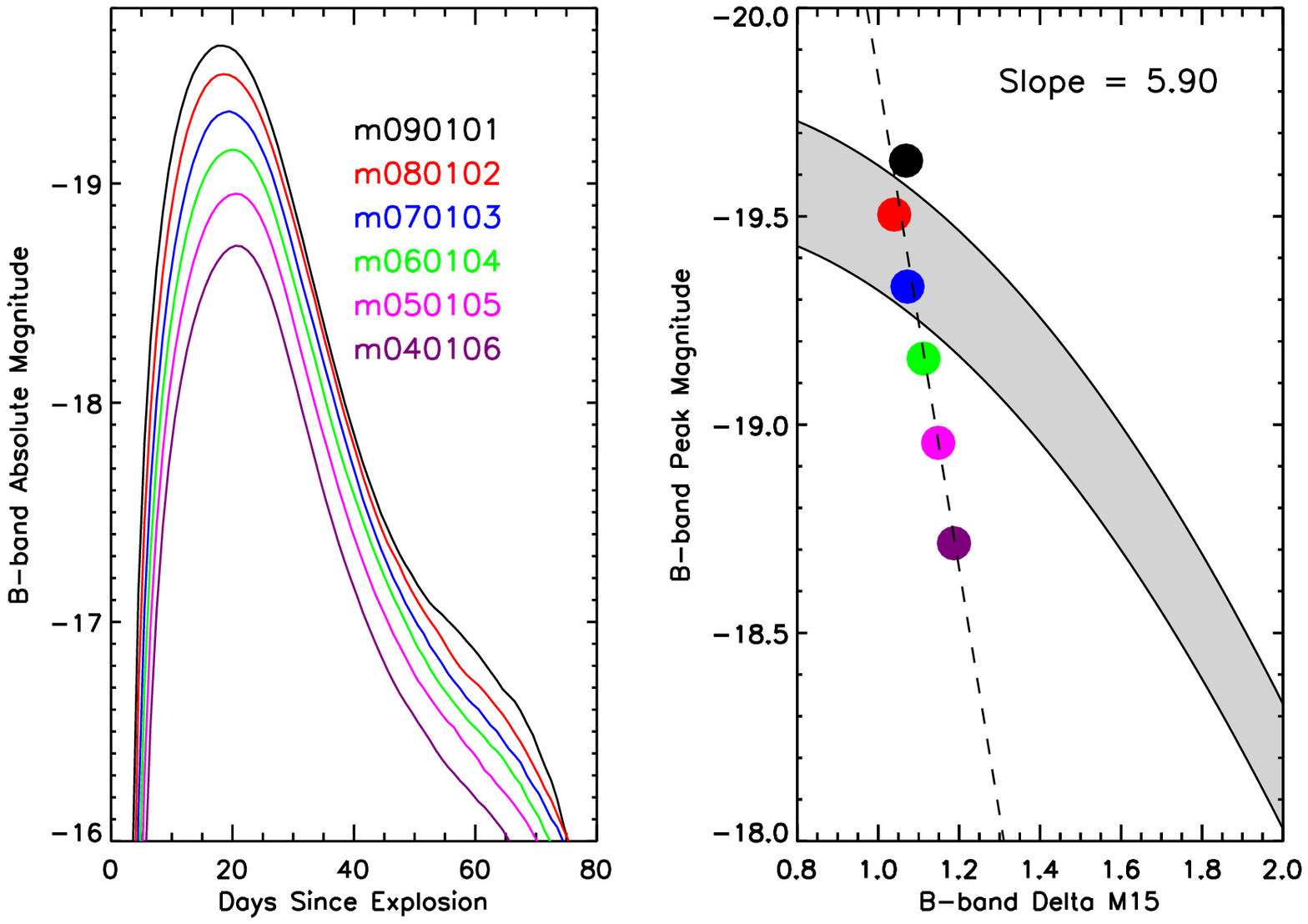}
\caption{Effect of varying \Mni\ at the expense of IME in moderately
  mixed models (Fig.~\ref{Fig:mixing}). The parameters are similar to
  Fig.~\ref{Fig:mbu_models}, except that the mass of stable iron is
  held fixed while the mass of IME varies: $\Mni =
  [0.4,0.5,0.6,0.7,0.8,0.9]$~\msun, $\Mfe = 0.1~\msun$, and $\Mime +
  \Mni + \Mfe = 1.1~\msun$.  The $B$-band peak magnitude declines even
  faster with decreasing \dmfb\ than the observations show.  This is
  because moderately mixed models with large masses of IME and low
  \Nifs\ have their \Nifs\ more centrally concentrated and thus
  decline more slowly. Better results are obtained with more mixing
  (Fig,\ref{Fig:nw_models}).
\label{Fig:ime_models}}
\end{center}
\end{figure}

\begin{figure}
\begin{center}
\includegraphics[clip=true,width=\columnwidth]{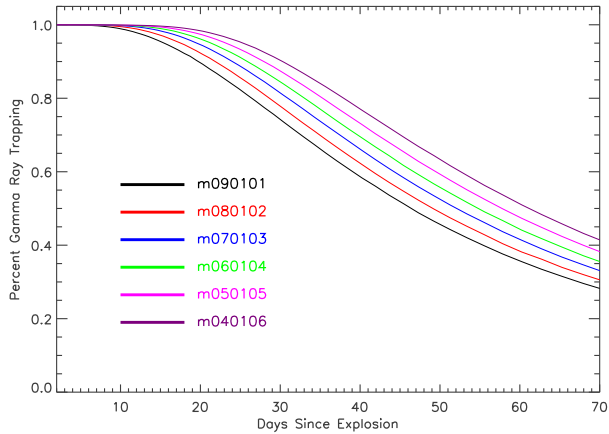}
\caption{Percentage of trapping of gamma-rays from radioactive decay
  as a function of time for the models of Fig.~\ref{Fig:ime_models} in
  which \Mni\ was varied.  Models with lower \Mni\ have
  \Nifs\ concentrated closer to the ejecta center, and hence more
  efficient gamma-ray trapping in the post-maximum epochs.
\label{Fig:gr_trapping}}
\end{center}
\end{figure}

Might the good agreement found in the last section persist if \Nifs\ is
varied at the expense, not of \Mfe, but of IME?  The models considered
here have fixed \Mfe = 0.1~\msun\ and \Mni\ varied from 0.4~\msun\ to
0.9~\msun\ at the expense of IME.  The procedure is similar to that
employed by \citet{Pin01}, but not identical because of issues of
mixing and, to a lesser extent, composition and explosion energy.  The
total iron group production here (\Mfe\ + \Mni) varied from 0.5 to
1.0~\msun\ while the burned mass was held fixed at \Mburn = 1.1~\msun.

Figure~\ref{Fig:ime_models} shows that the slope of the WLR for this
model set is steeper than observed.  The slight positive correlation
between \Mb\ and \dmfb\ is due to the dependence of the
spectroscopic/color evolution on \Mni, as discussed in \cite{Kas06} and
\Sect{basicwlr} above.  
However, three other effects act counter to this in the moderately
mixed models:

First, in the low \Mni\ models, the distribution of \Nifs\ is
concentrated closer to the ejecta center.  This leads to an increased
effectiveness in the trapping of gamma-rays from radioactive decay, as
shown in Fig.~\ref{Fig:gr_trapping}. Near maximum light ($\texp
\approx 20$~days) the differences in gamma-ray trapping among the
models are minor, but by day +15 ($\texp \approx 35$~days) trapping in
the \Mni = 0.3~\msun\ model is nearly 40\% greater than that in the
\Mni = 0.9~\msun\ model.  This extra trapping slows the decline rate
in the lower \Mni\ models.

Second, again because \Nifs\ is more highly concentrated towards the
center, the optical depth (and hence diffusion time) to the ejecta
surface is significantly larger in the low \Mni\ models (see
Eq.~\ref{Eq:lc_width}).  This further tends to slow the decline rate
of these models.

Finally, as mentioned several times already, the $B$-band decline rate
is determined largely by color evolution controlled by the development
of Fe~II and Co~II lines.  In the low \Mni\ models, the iron-group
elements extend to only very low velocities.  For example in the $\Mni
= 0.3$~\msun\ model, the edge of the \Nifs\ zone is at 5000~\kms\
compared to 8500~\kms\ for the \Mni = 0.7~\msun\ model.  This inhibits
the development of strong Fe~II/Co~II line blanketing in the low
\Mni\ models, and hence slows the decline rate of these models.

The very steep WLR seen in Fig.~\ref{Fig:ime_models} reflects the fact
that these three effects act counter to and nearly cancel the
spectroscopic, color-evolution effect.  Thus, in this particular set
of calculations, we find that simply varying the \Nifs\ mass at the
expense of IME in a well stratified medium does not give a WLR
relation with the correct slope.  Either this represents a failure of
the assumptions underlying the radiative transfer calculations, or (as
we discuss presently) at least a moderate degree of mixing of the
zones appears necessary.
 
\subsection{Centrally Concentrated Mixing}
\lSect{limmix}

\begin{figure}
\begin{center}
\includegraphics[clip=true,width=\columnwidth]{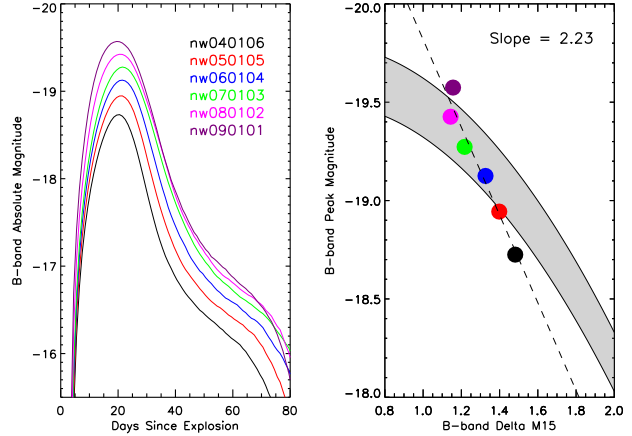}
\caption{Effect of varying \Mni\ at the expense of IME, but in explosions
  in which the inner 0.8 \Msun\ has been more thoroughly mixed (the
  ``MC'' Model in Fig.\ref{Fig:mixing}).  Unlike the stratified
  M-series models in Fig.~\ref{Fig:ime_models}, the MC-models better
  reproduce the observed WLR.
\label{Fig:nw_models}}
\end{center}
\end{figure}

\begin{figure}
\begin{center}
\includegraphics[clip=true,width=\columnwidth]{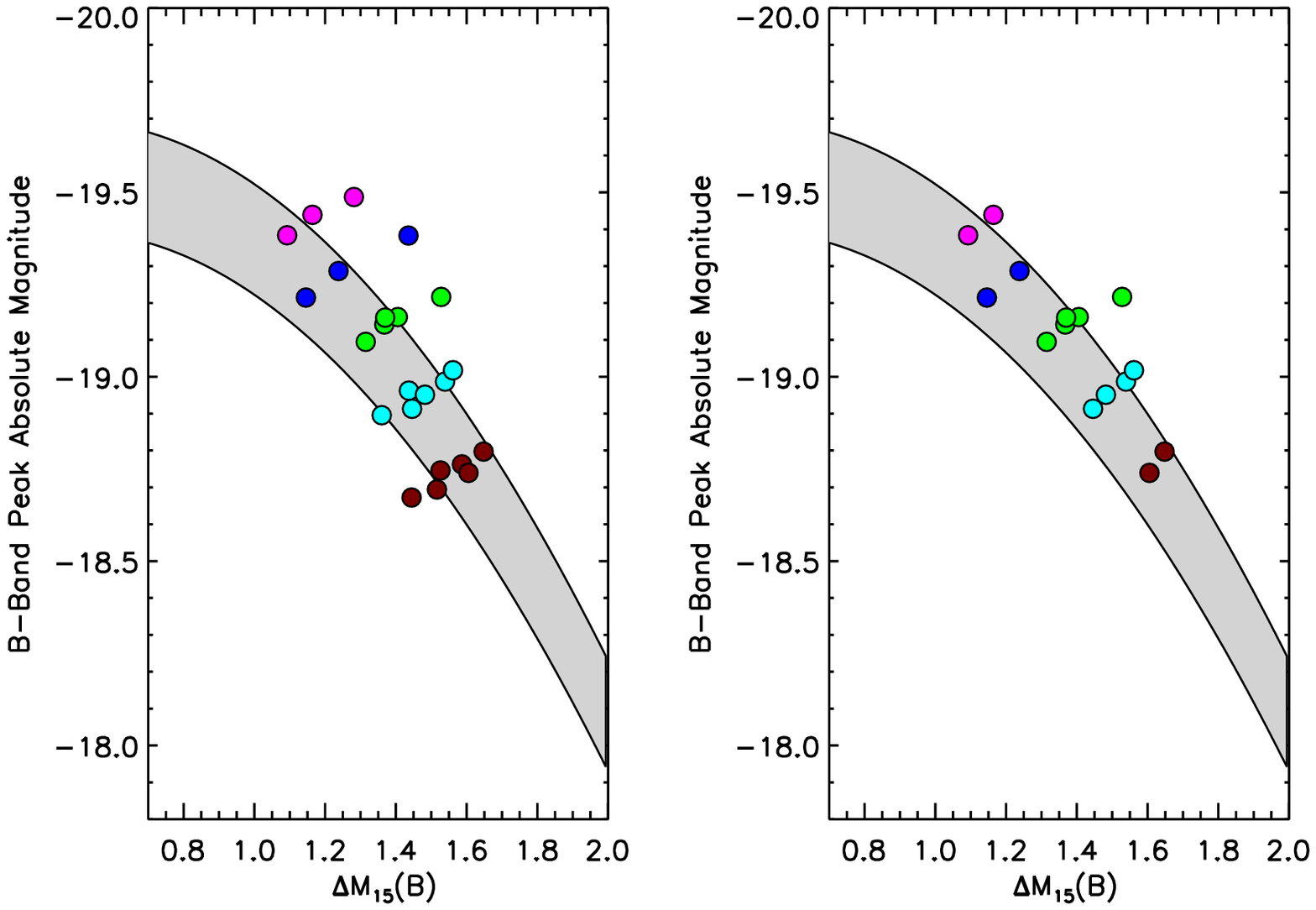}
\caption{\emph{Left:} WLR relation for the entire set of mixed core
  (MC-series) models.  These models are characterized by heavy mixing
  below $m = 0.8~\msun$.  As in Figure~\ref{Fig:m_finer}, they
  obey the constraints $\Mburn = 1.0-1.1~\msun$, $\Mfe =
  0.1-0.3$~\Msun, and $\Mime \ge 0.1$~\Msun.  The \Nifs\ mass varies
  from $\Mni = 0.4-0.8$~\Msun, with the same color coding as in
  Figure~\ref{Fig:all_ni}.  \emph{Right:} Same as left, but including
  only those models that have total iron group production between
  $\Mfe + \Mni = 0.7-0.9~\msun$.
\label{Fig:nw_all}}
\end{center}
\end{figure}

The models in \Sect{nivsime} failed to reproduce the observed WLR
because the distribution of iron group elements was more centrally
concentrated for the low \Mni\ models.  This can be rectified if the
the inner regions of ejecta are heavily mixed, thus equalizing the
distribution (but not quantity) of \Nifs\ among all models.  Such
mixing could, for example, reflect the maximum extent of the
deflagration region in a delayed detonation scenario. The outer part,
having been burned by a blast wave, would be less heterogeneous. If
the density were low enough, that part would be predominantly
IME. \footnote[2]{Detonation waves at intermediate densities ($\rho
  \sim {\rm few} \times 10^7$ g cm$^{-3}$) give explosive temperatures
  in the range $3.5 - 4.5 \times 10^9$ K where burning produces a
  mixture of \Nifs\ and IME. Lower densities give only IME plus neon,
  sodium, and magnesium. Eventually, the density is too low (and the
  heat capacity of the radiation field too high) for any burning.}

Observationally, the absence of significant high velocity iron in the
spectrum suggests that vigorous mixing was restricted to regions
$\ltaprx$0.8 \msun\ \citep{Branch_94D} . Transitions to detonation, if
they happen, are believed to occur when the burning enters the
``distributed regime'' at densities $\sim 1 - 3 \times 10^7$ g
cm$^{-3}$ \citep{Nie97}. Depending upon the detailed flame model and
especially how the burning is ignited, such densities are reached when
the flame has moved through $\sim$1 \Msun\ \citep[e.g.,][]{Nom84}.
This motivates the studies of models where the mixing has not been
uniform, but is concentrated in approximately the inner 0.8 \msun\ of
the star where it was assumed to be severe.  The MC-series of models
was prepared for a restricted range of burned mass, 1.0 and 1.1 \msun;
\Nifs\ masses, 0.4 to 0.9 \msun\; and stable iron masses, 0 - 0.3
\msun\ and a finer sampling of the intervals was also used. These
models had the same final elemental yields and explosion energies as
the corresponding models in the standard ``M'' series, but the
composition was distributed in a different way. These are called the
``MC'' (for ``mixed core'') models. Figure~\ref{Fig:mixing} shows an
example that should be compared with the corresponding ``M'' model.

In Fig.~\ref{Fig:nw_models} we show the equivalent of the models in
Fig.~\ref{Fig:ime_models} but for the MC series.  The mixing of the
inner layers leads to an increased agreement with the observed WLR.
Meanwhile, Fig.~\ref{Fig:nw_all} (left panel) shows the full set of
MC-series models.  Compared to the equivalent set of M-series models
(Fig.~\ref{Fig:m_finer}, left panel) the mixing serves to
significantly reduce the scatter and improve the agreement with the
observed WLR.  A more restrictive set of the MC models in which Fe +
\Nifs\ = 0.7 to 0.9 \msun\ fits the WLR even better
(Fig.~\ref{Fig:nw_all}, right panel).

This final plot shows that it is possible to define a well defined,
physically motivated cut of the model templates that does replicate
the observed WLR. The chief requisites are: 1) a nearly
constant explosion energy (and burned mass) around 1.0 - 1.2 \Msun; 2)
\Nifs\ varied at the expense of {\sl either} stable iron or IME, but
sufficiently well mixed in the inner 0.8 \Msun\ that its distribution
in velocity space does not vary greatly, even when the composition is
changed; and 3) a nearly constant iron plus \Nifs\ mass around $0.8\pm
0.1$ \Msun. The resulting WLR is still quite broad though, filling the
observed band with no excess left for observational errors and
systematics, and there is no {\sl a priori} reason why the explosions
must always be so constrained.

\section{CONCLUSIONS}

The multi-band photometry (and, in some cases, the spectra) of a large
set of Type Ia supernova models have been studied using two approaches
to the radiation transport problem. These models were constructed in a simple
fashion (\Sect{toys}) that allowed complete control over the major
parameters affecting the outcome - the explosion energy and the masses of
\Nifs, stable iron, and IME produced. The models were exploded by
depositing, uniformly, an amount of energy corresponding to the change
in composition in a standard white dwarf model. All sensitivity to
the uncertain physics of the actual burning was thus absorbed into the
three model parameters, plus some prescription for mixing. Comparison
of the light curves and spectra of sample models with those of more
complicated models and to observations of SN Ia gave good agreement.

Using these models, which consisted essentially of every explosion one
could produce starting from a 1.38 \msun \ carbon-oxygen white dwarf,
the necessary conditions were determined for reproducing the observed
relation between decline rate and peak magnitude (both in the
$B$-band). The physical parameters leading to intrinsic dispersion in
that relation were also identified and quantified.  The relation
between peak magnitude and color at peak magnitude was studied as
well, and found to be more robust to parameter variation.

The set of all explosions with positive kinetic energy did {\sl not}
give a WLR. Instead a large range of decline rates,
quite inconsistent with observations, was found for SN Ia with the
same mass of \Nifs. If SN Ia light curves are a single parameter
family, that parameter cannot be as simple as just the mass of \Nifs.
Other possible cuts of the model set were thus considered based upon
total kinetic energy (or mass burned), and various restrictions on the
final composition and mixing.

In order to satisfy the WLR, the ejecta structures of
SNe~Ia must obey certain constraints.  First, all SNe~Ia must have a
common total mass burned (and hence kinetic energy). In the present
calculations, good agreement with observations was found if that
common burned mass was $1.1 \pm 0.1$
\msun\ and the kinetic energy at infinity was $1.2 \pm 0.2 \times
10^{51}$ erg.  Second, the radial distribution of iron group
elements (including \Nifs) in the ejecta must be fairly uniform among
all SNe~Ia.  We found the best agreement with the observed WLR among
those models where the distribution of iron group elements extended
from near the center of the ejecta to $\sim$0.8
\msun.  For example, one can consider the subset of models in which 
 the total mass of iron-group elements is nearly constant at $\sim 0.8
\pm 0.1$~\Msun, with variations in \Mni\ arising from
difference in the ratio of \Nifs\ to stable iron group species
produced in the explosion.  Differences in this ratio are indeed
predicted to arise from metallicity and electron-capture effects,
although it is not clear that the entire range of SN~Ia luminosities
can be so explained given the constraints provided by the underlying
explosion physics and Galactic nucleosynthetic measurements.  Various
arguments lead to the conclusion that the mass of stable iron (mostly
$^{54,56}$Fe and $^{58,60}$Ni) is generally restricted to the range
0.1 to 0.3 \msun \ with about half of that range due to metallicity
effects.  Alternatively, one could vary the amount of \Nifs\ at the
expense of IME.  In this case, good agreement with the observed WLR is
found as long as the inner layers of ejecta ($\la 0.8$~\Msun) are
fairly well mixed. In the present calculations, well-stratified ejecta
structures do not reproduce the observed slope of the WLR, as the
lower \Mni\ models have \Nifs\ concentrated relatively closer to the
ejecta center, which tends to slow their light curve decline rate. To
populate the entire observed span of the WLR, especially
at faint luminosities, the mass of these IME is quite large in some
instances, arguing for a delayed detonation.

The present calculations adopted an LTE approximation for both the
level populations and the line source functions (i.e., the
thermalization parameter, $\epsilon_{\rm th}$ = 1).  Although a common
approach in many previous transfer studies, this (and other
simplifying approximations adopted) can be expected to lead to errors
in the model observables on the order of 0.1 to 0.3 mag.  The relative
differences between models, however, may be known more accurately, and
so the general trends obeyed by sets of models and the level of {\sl
dispersion} in such trends may be considered more robust.  For
example, our choice $\epth = 1$ overestimates the true redistribution
probabilities and leads to maximum light colors that are too red by
$\sim 0.06$~mag relative to the observations in the
color-peak-magnitude plot of, e.g., Fig.~\ref{Fig:bmv_constrain}.
Nevertheless, the clustering of the points in this plot is quite tight
and the slope is correct.  The possible refined calibration of the
radiative transfer calculations and its impact on the study of the WLR
will be explored further in a subsequent paper.

Given the large dispersion found, even in selected subsets of our
models (e.g., Fig.~\ref{Fig:all_constrain}), the narrow spread in the
observed relation between \Mb\ and \dmfb\ is surprising. No single
physical parameter yields it, unless several other parameters are
highly constrained. Part of that spread can be reduced by a judicious
choice of mixing and a tight constraint on the mass burned
(Fig.~\ref{Fig:nw_all}). Even so, it is difficult to avoid the
conclusion that a major fraction of the observed scatter of the WLR,
$\gtaprx 0.1$~mag, reflects intrinsic physical diversity, and not
observational effects. This conclusion has important implications as
one plans for future studies using SN Ia as calibrated standard
candles to ever higher precision.  Given the dispersion expected from
the models themselves, a larger sample of supernovae will have to be
studied to get a strong signal to noise ratio for cosmological
effects.

A primary benefit of the approach adopted in this paper is the ability
to construct an expansive grid of SNe~Ia explosion models without
restricting ourselves to any specific theoretical explosion paradigm.
Such a large and general database of model light curves and spectra
nicely compliments the large sample of SNe~Ia data sets currently
being acquired by ongoing observational programs.  A direct comparison
of the models to individual observations would be useful in
interpreting the physical properties of any given SN~Ia event.  In
addition, the model grid should be helpful in developing refined
techniques for calibrating the luminosities of SNe~Ia so as to limit
and reduce intrinsic sources of dispersion or evolution.  For example,
our studies suggest that an improved strategy might be {\sl
spectroscopic} template fitting in which as much data as possible from
each individual SN Ia is compared to a template of model spectra and
colors. The models employed could be the ones computed here or some
derivative of that set.

\begin{acknowledgements}

SEW acknowledges support from NASA (NNG05GG08G) and the DOE Program
for Scientific Discovery through Advanced Computing (SciDAC;
DE-FC02-01ER41176).  DNK is supported by the Allan C. Davis fellowship
at Johns Hopkins University and the Space Telescope Science Institute.
SIB is supported by MPA guest program, and thanks especially
W.Hillebrandt and F.R\"opke for cooperation.  In Russia, this work is
supported also partly by the Russian Foundation for Basic Research
(projects 05-02-17480, 04-02-16793), by grants of `Scientific Schools'
of Russian Ministry of Science, and by grant IB7320-110996/1 of the
Swiss National Science Foundation.  This research used resources of
the National Energy Research Scientific Computing Center, which is
supported by the Office of Science of the U.S. Department of Energy
under Contract No.  DE-AC03-76SF00098.

\end{acknowledgements}

\end{document}